\documentclass[draftcls,onecolumn]{IEEEtran}

\usepackage{nicefrac}
\usepackage[utf8]{inputenc}
\usepackage[T1]{fontenc}
\usepackage{url}
\usepackage{ifthen}
\usepackage{cite}
\usepackage[cmex10]{amsmath}
\usepackage{graphicx}
\usepackage{amsmath, amsthm, amssymb, amsfonts}
\usepackage{tabularx}
\usepackage[linesnumbered,ruled,vlined]{algorithm2e}
\usepackage{subfigure}
\usepackage{url}
\usepackage{xcolor}
\usepackage{multirow}
\usepackage{colortbl}
\newtheorem{definition}{Definition}
\newtheorem{lemma}{Lemma}
\newtheorem{corollary}{Corollary}
\newtheorem{theorem}{Theorem}
\newtheorem{remark}{Remark}

\newtheorem{construction}{Construction}
\usepackage[hidelinks]{hyperref}
\usepackage{enumitem}
\usepackage{cleveref}
\crefname{algorithm}{Alg.}{Algs.}
\crefname{AlgoLine}{line}{lines}
\newcounter{algoline}

\renewcommand{\nllabel}[1]{\refstepcounter{algoline}\label{#1}}
\crefname{algoline}{line}{lines}
\Crefname{algoline}{Line}{Lines}
\newcommand{\lb}[1]{\ensuremath{\mathsf{lb}(#1)}}
\newcommand{\ub}[1]{\ensuremath{\mathsf{ub}(#1)}}

\begin{document}
\title{Optimal Universal Coding of Integers}

\author{Wei Yan,
        Yunghsiang S. Han,~\IEEEmembership{Fellow,~IEEE},
        and Leqian Zheng
\thanks{
W. Yan and L. Zheng are with National University of Defense Technology, China, email: yan.wei2023@nudt.edu.cn, zheng.leqian@nudt.edu.cn.}
\thanks{Y. S. Han is with Great Bay University, China. Email: yunghsiangh@gmail.com.
}
}
\maketitle
\begin{abstract}
Universal coding of integers (UCI) provides prefix codes for positive integers such that, for every nonincreasing source distribution $P$, 
the average codeword length under $P$ stays within $K$ times $\max\{1,H(P)\}$. 
For a given UCI $\mathcal{C}$, the smallest such constant $K$ is called the minimum expansion factor of $\mathcal{C}$, denoted $C_{\mathcal{C}}^{*}$.
Let $C^{*}\triangleq \inf \{C_{\mathcal{C}}^{*}\}$ over all UCIs. 
A UCI attaining $C^{*}$, or a family of UCIs attaining $C^{*}$ in the limit, is called optimal. 
It is currently known that $2\le C^{*}\le 2.0386$. 
In this paper, we construct a family of one-point plus uniform-tail distributions and prove that, for every UCI, a distribution in this family attains the worst-case ratio; thus, the family is least favorable for the UCI problem.
We further establish an inequality, called the
\emph{UCI inequality}, which plays the same role for UCI as the Kraft inequality does for prefix codes: for any real number $B\ge 1$, it decides whether $B$ lies below or above $C^*$. The UCI inequality gives an equivalent definition of $C^*$. By numerical computation, we determine $C^* = 2.00012475703610197861701176297579886254\cdots$, the first $38$ decimal digits being certified. Once $C^*$ is known, we can theoretically construct the optimal UCI.
\end{abstract}
\IEEEpeerreviewmaketitle

\section{Introduction}\label{sec_intro}
Source coding~\cite{SH481} removes the redundancy of a source so that the average number of bits per source symbol approaches the entropy, 
which in turn raises the effectiveness and the transmission efficiency of communication systems.
To this end, one employs a variable-length code~(VLC): each source symbol is assigned a binary word whose length may vary from symbol to symbol, with frequent symbols receiving short words and rare symbols long ones.
When the source distribution $P$ is known, this design task is completely settled---Huffman's algorithm~\cite{H52} yields prefix codes whose expected length lies between $H(P)$ and $H(P)+1$~\cite{EIT}.

In practical scenarios where the probability distribution of a discrete memoryless source~(DMS) is unknown, Elias~\cite{Elias75} introduced the Universal Coding of Integers~(UCI), an important class of VLCs for such scenarios.
The alphabet that UCI is designed for is countably infinite---the positive integers---and the code is required to work without assuming any
statistical knowledge of the source.
UCI has been widely applied in various fields, including evolving secret sharing~\cite{YLH23,Cheng25}, quantized stochastic gradient descent~\cite{NIPS}, stream processing engines~\cite{Zhang23,Zhang24}, and data deduplication schemes~\cite{N19,LF22}.

The universal coding problem that Elias~\cite{Elias75} considers is as follows.
The source is a DMS whose probability distribution $P$ is defined on the alphabet $\mathbb N=\{1,2,\dots\}$ 
and satisfies $P(1)\ge P(2)\ge\cdots$ and $\sum_{m=1}^\infty P(m)=1$.
Such a distribution \(P\) is nonincreasing.
The entropy \(H(P)\) of \(P\) is defined as $H(P)\triangleq -\sum_{m=1}^\infty P(m)\log_2 P(m)$. 
Let $\mathcal{P}_\mathbb{N}$ denote the set of all nonincreasing distributions on $\mathbb N$ with $H(P)<\infty$.
A code $\mathcal{C}$ maps each integer $m$ to a finite binary word; 
the number of bits in that word is denoted $L_\mathcal{C}(m)$, and the code is required to be prefix-free, 
which is equivalent to the condition $\sum_{m=1}^\infty 2^{-L_\mathcal{C}(m)}\le 1$~\cite{kraft}. 
The average codeword length under $P$ is defined as $A_P(L_\mathcal{C})\triangleq \sum_{m=1}^\infty  L_\mathcal{C}(m)P(m)$.
Elias~\cite{Elias75} defined this code to be \emph{universal}, in the sense that it must satisfy the following requirements:
the average codeword length must remain within a constant factor $K$ of $\max\{1,H(P)\}$ no matter which \(P \in \mathcal{P}_\mathbb{N}\) is chosen, i.e.,
\begin{equation}\label{eq50}
A_P(L_\mathcal{C})\;\le\;K\max\{1,H(P)\}
\end{equation}
for all $P \in \mathcal{P}_\mathbb{N}$. The smallest constant $K$ satisfying Equation~\eqref{eq50} is called the \emph{minimum expansion factor} of $\mathcal{C}$, denoted $C_{\mathcal{C}}^{*}$~\cite{YL21}, i.e.,
\begin{equation*}
\begin{aligned}
     C_{\mathcal{C}}^{*}&\triangleq  \inf \left\{ K \ \Big|\ A_P(L_\mathcal{C})\;\le\;K\max\{1,H(P)\} \mbox{ for }\forall P \in \mathcal{P}_\mathbb{N} \right\}       \\
                        &= \sup_{P\in\mathcal{P}_\mathbb{N}} \frac{A_{P}(L_{\mathcal{C}})}{\max\{1,H(P)\}} \ .
\end{aligned}
\end{equation*}
Let $S_{UCI}$ denote the set of all UCIs.
The smallest $C_{\mathcal{C}}^{*}$ is called \emph{optimal minimum expansion factor}, denoted $C^*$~\cite{Yan26}, i.e.,
\[
\begin{aligned}
     C^{*}&\triangleq \inf_{\mathcal{C}\in S_{UCI}}  C_{\mathcal{C}}^{*} \\
          &= \inf_{\mathcal{C}\in S_{UCI}} \sup_{P\in\mathcal{P}_\mathcal{A}} \frac{A_{P}(L_{\mathcal{C}})}{\max\{1,H(P)\}} \ .
\end{aligned}
\]
A prefix code or a family of prefix codes,\footnote{For a family of UCIs $\{\mathcal{C}_m\}_{m=1}^{\infty}$,
its minimum expansion factor is understood as $\lim\limits_{m\rightarrow\infty}C_{\mathcal{C}_m}^*$.} that achieves $C^*$ is called an \emph{optimal UCI}.

The early work on UCI can be broadly divided into two categories~\cite{C1990,AS17}.
\begin{enumerate}
\item \emph{Message length scheme}: This class of coding schemes encodes a positive integer \(m\) into two parts: a suffix that uses \(s\) bits to represent \(m\), and a prefix that encodes \(s\).
For most codes of this scheme, the design goal is to keep the codeword length $L_{\mathcal{C}}(m)$ as short as possible for large $m$. This scheme was proposed in~\cite{Elias75,L68,ER78,S80,Y00}.
\item \emph{Flag scheme}: This class of coding schemes uses a special string as an end-of-codeword flag.
The flag scheme enjoys a self-synchronization property in decoding, which stops errors from propagating without bound. This scheme was proposed in~\cite{L81,AF87,W88,YO91,AS17}.
\end{enumerate}
Recently, Yan \emph{et al.}~\cite{YL21,YL22,Yan26} have been working on narrowing the range of the optimal minimum expansion factor \(C^*\), to find the optimal UCI.
In 1975, Elias's work~\cite{Elias75} showed that the range of \(C^*\) is \(1\le C^*\le 3\), where the \(\gamma\) code achieves \(C_{\gamma}^{*}\le 3\).
It was established in~\cite{YL21,YL22} that the range of
$C^{*}$ was narrowed to the interval $[2,\,2.5]$.
The lower bound was raised from \(1\) to \(2\) mainly because a class of distributions \(P_n\)~\cite{YL21} was constructed for an arbitrary UCI code \(\mathcal{C}\), regardless of how \(\mathcal{C}\) is constructed; the construction is as follows.
\begin{equation} \label{eq51}
P_{n}(m)=\left\{\begin{array}{lll}
	1-\frac{1}{n},            &\text{if } m=1\text{ ,}\\
	\frac{1}{n2^n},    &\text{if } m=2,3,\ldots,2^n+1 \text{ ,}\\
	0,                          &\text{otherwise.}   \\
\end{array}\right.
\end{equation}
The upper bound was tightened from \(3\) to \(2.5\), mainly because the Wyner’s
inequality~\cite{Wyner72} 
\[
 \sum_{m=2}^\infty P(m)\log_2 m \le H(P)
\]
used by Elias~\cite{Elias75} was improved to
\[
 \sum_{m=2}^\infty P(m)\log_2 m \le P(1)\log_2 P(1)+H(P),
\]
and the corresponding $\iota$ code~\cite{YL22} was constructed.
The work in~\cite{Yan26} further improved the inequality required for proving the upper bound; the inequality was improved to
\[
\sum_{m=2}^{\infty} P(m) \log_2 m \le H(P) - h(P(1)) + P(2) 
+ \sum_{m=2}^{\infty} P(m+1) \log_2 \left( 1 + \frac{1}{m} \right) \left( 1 - \frac{1}{m} \right)^{m-1},
\]
where $h(x)\triangleq-x\log_{2}x-(1-x)\log_{2}(1-x)$ denotes the binary entropy of $(x,1-x)$, and the corresponding $\nu$ code was constructed such that $C_\nu^*\leq 2.0386$. Therefore, the current range of the optimal minimum expansion factor is \(2\le C^* \le 2.0386\). A generalized version of universal coding of integers, closely connected to UCIs, has also been introduced and studied in~\cite{ITW,YH24,Yan25}.

In this paper, we further narrow the upper and lower bounds on $C^*$ and, in fact, pin down $C^*$ itself to $38$ certified decimal digits.  
Since the proof in~\cite{Yan26} for tightening the upper bound on \(C^*\) by improving inequalities has already been divided into $13$ cases, we do not pursue this direction further.
Inspired by the distribution $P_n$ shown in Equation~\eqref{eq51}, 
we consider constructing \emph{least favorable distributions} for any UCI \(\mathcal{C}\in S_{UCI}\). 
In the UCI problem, a family of distributions is said to be least favorable if, for any UCI $\mathcal{C}\in S_{UCI}$, the worst-case ratio $A_P(L_{\mathcal{C}})/\max\{1,H(P)\}$ over all admissible $P$ is attained by a member of this family; the attained value is then exactly $C_{\mathcal{C}}^*$.
Finally, we prove that $C^* = 2.000124757036101\cdots$ and provide the corresponding construction.
We summarize the main contributions of this paper as follows.
\begin{enumerate}
\item Inspired by Equation~\eqref{eq51}, 
we construct a family of distributions called the
\emph{one-point plus uniform-tail distributions}, 
and prove that, for all $\mathcal{C}\in S_{UCI}$, 
the worst-case ratio $A_P(L_{\mathcal{C}})/\max\{1,H(P)\}$ is attained by a distribution in this family. 
These distributions then serve as the starting point for deriving upper and lower bounds for $C^*$.
\item In our study of UCI, we establish an important inequality that plays a role analogous to that of the Kraft inequality in the study of prefix codes; we call it the \emph{UCI inequality}.
For any real number $B\ge 1$, checking the UCI inequality decides at once whether \(B\) is below or above \(C^*\) 
\item Using the UCI inequality, we obtain an equivalent definition of $C^*$. By numerical computation, we determine $C^* = 2.00012475703610197861701176297579886254\cdots$, whose first $38$ decimal digits are certified. Moreover, once \(C^*\) is determined, we have theoretically constructed an optimal UCI.
\end{enumerate}

The rest of this paper is organized as follows.
Section \ref{pre} introduces the notation and reviews relevant results on the minimum expansion factor of UCI.
Section \ref{new} defines the one-point plus uniform-tail distributions and proves that they are the least favorable for the UCI problem.
In Section \ref{Inequality}, we introduce and prove the UCI inequality, characterize \(C^*\) equivalently by means of it, and construct the optimal UCI in theory.
Section \ref{sec_con} summarizes this work.

\section{Preliminaries}\label{pre}
In this section, we introduce the notation used in this paper and present relevant results on the minimum expansion factor of UCI. 
Table~\ref{tab3} summarizes the main notations used in this paper.

\begin{table}[!ht]
\centering
\caption{Main notations used in this paper}\label{tab3}
\begin{tabular}{c|l|c}
\hline \hline
Notation  & Description  & Definition \\
\hline \hline
$P$   &     The nonincreasing probability distribution &  Section \ref{sec_intro}     \\
\hline
$\mathbb N$    & The countably infinite alphabet or the set of positive integers   &   Section \ref{sec_intro}                 \\
\hline
$H(P)$   &    The entropy of $P$  & Section \ref{sec_intro}      \\
\hline
$\mathcal{P}_\mathbb{N}$  & the set of all nonincreasing distributions on $\mathbb N$ with $H(P)<\infty$   &  Section \ref{sec_intro}         \\
\hline
$\mathcal{C}$   &   The prefix code  &  Section \ref{sec_intro}         \\
\hline
$L_{\mathcal{C}}(m)$  &   The codeword length of \(m\) under code $\mathcal{C}$
& Section \ref{sec_intro}     \\
\hline
$A_P(L_{\mathcal{C}})$   & The average codeword length of $\mathcal{C}$ under $P$ & Section \ref{sec_intro}   \\
\hline
$C_{\mathcal{C}}^*$  & The minimum expansion factor of UCI $\mathcal{C}$    &  Section \ref{sec_intro}         \\
\hline
$S_{UCI}$           &  The set of all UCIs    &  Section \ref{sec_intro}         \\
\hline
$C^*$                 &   The optimal minimum expansion factor    &   Section \ref{sec_intro}      \\
\hline
$h(x)$               &   The binary entropy of $(x,1-x)$         &   Section \ref{sec_intro}      \\
\hline
$U_k$   &  The uniform distribution on $\{1,2,\ldots,k\}$   &    Section \ref{pre}          \\
\hline
$a_k$     &    The average codeword length under $U_k$      &   Section \ref{pre}    \\
\hline
$\widetilde{U}_k$   &  The uniform distribution on $\{2,3,\ldots,k\}$   &    Section \ref{pre}          \\
\hline
$\widetilde{a}_k$     &    The average codeword length under $\widetilde{U}_k$      &   Section \ref{pre}    \\
\hline
$D(Q\|P)$                  &   The relative entropy between two probability distributions \(Q\) and \(P\)    &    Eq.~\eqref{eq52}                   \\
\hline
$X\sim P$     & The random variable $X$ is distributed according to $P$     &   Section \ref{pre}    \\
\hline
$H(X)$       &  $H(X)=H(P)$ whenever $X\sim P$     &   Section \ref{pre}    \\
\hline
$H(X\mid Y)$   &  The conditional entropy of $X$ given $Y$    &  Eq.~\eqref{eq53}   \\
$I(X;Y)$   &  The mutual information between $X$ and $Y$ &    \\
\hline
$P_{\varepsilon,k}$    &   The one-point plus uniform-tail distribution with parameters \(\varepsilon\) and \(k\)  &   Eq.~\eqref{def One-point}  \\
\hline
$\mathcal{P}(\varepsilon,k)$    &   The set of all one-point plus uniform-tail distributions  &   Eq.~\eqref{set One-point}  \\
\hline
$\varepsilon_x$   & The unique solution to $h(t)+t\log_{2}(x-1)=1$ on $(0,\frac{x-1}{x}]$   &    Section \ref{new}          \\
\hline
$\sigma_x$   &  $\sigma_x=\varepsilon_x^{-1}$ & Section \ref{new}     \\
\hline
$R_k(\varepsilon)$   &  $R_k(\varepsilon)=\dfrac{A_{P_{\varepsilon,k}}(L)}{\max\{1,H(P_{\varepsilon,k})\}}$ & Eq.~\eqref{notation1}     \\
$\overline{B}$   &  $	\overline{B}=\sup_{P\in\mathcal{P}(\varepsilon,k)}\dfrac{A_{P}(L)}{\max\{1,H(P)\}}            $ &     \\
\hline
$\Phi(y)$   &  $\Phi(y)=y\left(1-h\left(\frac{1}{y}\right)\right),\qquad y\ge 2$ & Eq.~\eqref{function}     \\
\hline
$Y$  &  \(Y\) is the inverse function of \(\Phi\) & Eq.~\eqref{function Y}   \\
\hline
$T(\mathcal{C})$  &  \(T(\mathcal{C})=\sup_{2\leq k\in\mathbb{N}}  \max \left\{(1-\varepsilon_k)L_{\mathcal{C}}(1)+\varepsilon_k\widetilde{a}_k, \frac{a_k}{\log_2 k} \right\}\)  & Eq.~\eqref{eq54}   \\
\hline
$S(k)$  &  The sum of the lengths of the first $k$ codewords  & Section \ref{Inequality}   \\
\hline
$N_n$  &  The number of codewords of length $n$  & Section \ref{Inequality}   \\
\hline
$\mathcal{A}_n$  &  The number of codewords of length at most \(n\)  & Section \ref{Inequality}   \\
\hline
$\mathcal{B}_n$  &  \(\mathcal{B}_n=\sum_{j=1}^{n} \mathcal{A}_j\)  & Section \ref{Inequality}   \\
\hline
$c_{B,\mathcal{C}}(k)$   &  $c_{B,\mathcal{C}}(k)=\min \Big\{L_{\mathcal{C}}(1)+(k-1)\big[L_{\mathcal{C}}(1)+\big(B-L_{\mathcal{C}}(1)\big)\sigma_k\big],Bk\log_2 k\Big\}$ & Eq.~\eqref{eq55}     \\
$d_{n,\mathcal{C}}(B) $   &  $d_{n,\mathcal{C}}(B) =\sup_{2\le k \in \mathbb{N}} \{nk- c_{B,\mathcal{C}}(k)\}       $ &     \\
\hline
$c_{B}(k)$   &  $c_{B}(k)= \min \Big\{1+(k-1)\big[1+(B-1)\sigma_k\big],Bk\log_2 k\Big\}$ & Eq.~\eqref{eq56}     \\
$	d_{n}(B) $   &  $d_{n}(B)= \sup_{2\le k \in \mathbb{N}} \{nk- c_{B}(k)\}    $ &     \\
\hline
$\Sigma(B)$   &  $\Sigma(B)= \sum_{n=3}^\infty \lceil d_{n}(B) \rceil 2^{-n}$ & Eq.~\eqref{Sigma}     \\
\hline \hline
\end{tabular}
\end{table}

\subsection{Notations}
In this subsection, we first introduce some commonly used notation.
Let $U_k$ denote the uniform distribution on $\{1,2,\ldots,k\}$, so that $H(U_k)=\log_{2}k$.
Let $a_k\triangleq\frac{1}{k}\sum_{n=1}^{k}L(n)$ denote the average codeword length under the distribution $U_k$.
Let $\widetilde{U}_k$ denote the uniform distribution on $\{2,3,\ldots,k\}$,
then the entropy $H(\widetilde{U}_k)=\log_{2}(k-1)$.
Let $\widetilde{a}_k\triangleq\frac{\sum_{n=2}^{k} L(n)}{k-1}$ denote the average codeword length under the distribution $\widetilde{U}_k$.
The relative entropy between two probability distributions \(Q\) and \(P\) is defined as
\begin{equation} \label{eq52}
D(Q\|P)\triangleq\sum_{n\in\mathbb{N}} Q(n) \log_2 \frac{Q(n)}{P(n)}.
\end{equation}
Let $X$ be a random variable on the alphabet $\mathbb{N}$.
The notation $X\sim P$ means that $X$ is distributed according to the probability distribution $P$,
i.e., $\Pr(X=n)=P(n)$ for all $n\in\mathbb{N}$.
We extend the standard notation for entropy to random variables in the obvious way: $H(X)\triangleq H(P)$ whenever $X\sim P$.
Let $X$ and $Y$ be two random variables on the alphabet $\mathbb{N}$.
The conditional entropy $H(X\mid Y)$ and the mutual information $I(X;Y)$ are defined as follows.
\begin{equation} \label{eq53}
\begin{aligned}
			H(X\mid Y)&\triangleq \sum_{n=1}^\infty \Pr(Y=n)H(X\mid Y=n)   \\
			      &=-\sum_{n=1}^\infty \Pr(Y=n)\sum_{m=1}^\infty \Pr(X=m\mid Y=n)\log_{2} \Pr(X=m\mid Y=n), \\
           I(X;Y)&\triangleq \sum_{m=1}^\infty \sum_{n=1}^\infty \Pr(X=m,Y=n)\log_{2} \dfrac{\Pr(X=m,Y=n)}{\Pr(X=m)\Pr(Y=n)}.        \\
\end{aligned}
\end{equation}

The following theorem provides the classical information-theoretic result used in this paper.
\begin{theorem}\cite{EIT} \label{thm2}
\begin{enumerate}
\item $D(Q\|P)\geq 0$ for any probability distributions \(Q\) and \(P\);
\item $H(X)=H(X\mid Y)+I(X;Y)$ for any random variables $X$ and $Y$.
\end{enumerate}
\end{theorem}

\subsection{Conventions and some results for UCI}
In this subsection, we first highlight a convention in UCI and provide a brief note on the notation used in this paper.

Since the probability distributions \(P\) considered in the UCI problem are nonincreasing, if the average codeword length is to be minimized, the prefix code \(\mathcal{C}\) must satisfy
\begin{equation}  \label{eq60}
L_\mathcal{C}(m) \le L_\mathcal{C}(m+1)
\end{equation}
for all \(m\in\mathbb{N}\). 
Elias~\cite{Elias75} called a code \(\mathcal{C}\) \emph{minimal} if it satisfies Equation~\eqref{eq60}.
We adopt the convention that all UCIs considered in this paper's proofs are minimal.
This convention is motivated by a requirement common to all results on optimal UCIs, namely that the first codeword has length $1$~\cite{YL22}. Without minimality, a later codeword could have length $1$ while the first does not, and some proofs would then need additional assumptions.
Restricting to minimal UCIs costs nothing, since any non-minimal UCI can be converted into a minimal one by permuting its codewords, and, because $P$ is nonincreasing, the average codeword length does not increase under such a permutation.

Moreover, the symbols \(L(m)\) and \(A_P(L)\) may appear in the subsequent proofs in this paper. Here, \(L(m)\) denotes the codeword length of $m$, and \(A_P(L)\)  denotes the average codeword length under $P$.
Only the subscript $\mathcal{C}$ identifying the code has been dropped: several results below hold for every prefix code $\mathcal{C}$, and the subscript is suppressed for notational simplicity.
Finally, we present two results concerning the minimum expansion factor of UCI.
\begin{theorem}~\cite{YL21,YL22}\label{thm1}
Let $\mathcal{C}$ be an arbitrary UCI.
Then its minimum expansion factor satisfies $C_{\mathcal{C}}^{*}\geq 1+L_{\mathcal{C}}(1)$, which further implies that
$C_{\mathcal{C}}^{*}\geq2$ holds for every UCI $\mathcal{C}$.
\end{theorem}

\begin{theorem}~\cite{Yan26}\label{thm9}
The optimal minimum expansion factor $C^*$ satisfies $2\le C^* \le 2.0386$, where $C_\nu^*\leq 2.0386$ is achieved by $\nu$ code.
\end{theorem}

\section{Least Favorable Distributions}\label{new}
This section introduces the one-point plus uniform-tail distributions and establishes their least favorability for the UCI problem.

\subsection{One-point plus uniform-tail distributions}
Let $k\ge2$ be an integer and let $\varepsilon\in(0,\frac{k-1}{k}]$ be a real number.
Define the one-point plus uniform-tail distribution $P_{\varepsilon,k}$ as follows.
\begin{equation}\label{def One-point}
P_{\varepsilon,k}(n)=\left\{\begin{array}{lll}
1-\varepsilon,            &\text{if } n=1\text{ ,}\\
\dfrac{\varepsilon}{k-1},    &\text{if } n=2,3,\ldots,k \text{ ,}\\
0,                          &\text{otherwise.}   \\
\end{array}\right.
\end{equation}
By the definition of $P_{\varepsilon,k}$,
the distribution has a point mass at $1$ and is uniform on the remaining support;
we therefore call it the one-point plus uniform-tail distribution.
Let $\mathcal{P}(\varepsilon,k)$ denote the set of all one-point plus uniform-tail distributions; that is,
\begin{equation}\label{set One-point}
\mathcal{P}(\varepsilon,k)=\left\{P_{\varepsilon,k}\,\Big|\,2\leq k\in\mathbb{N},0<\varepsilon\leq\frac{k-1}{k}\right\}.
\end{equation}
From $\varepsilon\leq\frac{k-1}{k}$, we obtain
\[
 P_{\varepsilon,k}(1)=1-\varepsilon\geq \dfrac{\varepsilon}{k-1}= P_{\varepsilon,k}(2),
\]
and hence $\mathcal{P}(\varepsilon,k)\subseteq\mathcal{P}_\mathbb{N}$.
Also note that the uniform distribution $U_k=P_{\frac{k-1}{k},k}\in \mathcal{P}(\varepsilon,k)$.
\textbf{T}he entropy is
\begin{equation*}
\begin{aligned}
H(P_{\varepsilon,k})
&= -(1-\varepsilon)\log_2(1-\varepsilon) - \frac{\varepsilon}{k-1} \cdot k-1 \cdot \log_2 \frac{\varepsilon}{k-1}  \\
&= h(\varepsilon) + \varepsilon \log_2(k-1).   \\
\end{aligned}
\end{equation*}
The average codeword length is
\begin{equation*}
\begin{aligned}
A_{P_{\varepsilon,k}}(L) &= (1-\varepsilon)L(1) + \frac{\varepsilon}{k-1} \sum_{n=2}^{k} L(n)  \\
                         & = (1-\varepsilon)L(1) + \varepsilon \widetilde{a}_k.
\end{aligned}
\end{equation*}

We now consider the special case $H(P_{\varepsilon,k})=1$, namely 
\begin{equation}\label{eq:H=1}
h(\varepsilon) + \varepsilon \log_2(k-1)=1.
\end{equation}
In Equation~\eqref{eq:H=1}, the integer $k$ enters only through the term $\log_2(k-1)$; in Section~\ref{Inequality},  the same expression has to be evaluated at a real parameter. We therefore solve Equation~\eqref{eq:H=1} once for a real parameter $x\ge2$, and specialize to integers afterward.

For a real number $x\ge2$, define the function
\[
f_x(t)\triangleq h(t)+t\log_2(x-1),\qquad t\in\Big(0,\frac{x-1}{x}\Big].
\]
Taking the derivative yields
\[
f_x'(t)=\log_2\frac{(x-1)(1-t)}{t},
\]
so $f_x'(t)\ge0$ if and only if $(x-1)(1-t)\ge t$, i.e., $t\le\frac{x-1}{x}$; hence $f_x(t)$ is strictly increasing on $\big(0,\frac{x-1}{x}\big]$. 
 Furthermore, since
\begin{equation*}
	\begin{aligned}
		\lim_{t\to0^+}f_x(t) & =\lim_{t\to0^+}h(t)+0 = 0, \\
	f_x\!\left(\frac{x-1}{x}\right) &  =h\!\left(\frac{x-1}{x}\right)+\frac{x-1}{x}\log_2(x-1)=\log_2x\ \ge\ 1,
	\end{aligned}
\end{equation*}
$f_x(t)=1$ has a unique solution in $\big(0,\frac{x-1}{x}\big]$; 
we denote this unique solution by $\varepsilon_x$.
Moreover, by the monotonicity of $f_x$ and
\[
f_x\!\left(\frac{1}{2}\right)=1+\frac{1}{2}\log_2(x-1)\ \ge\ 1,
\]
it follows that $0<\varepsilon_x\leq \frac{1}{2}$.
Define
\[
	\sigma_x\triangleq\frac{1}{\varepsilon_x}\in[2,\infty),
\]
equivalently,
\begin{equation}\label{eq:sigma}
	h\left(\frac{1}{\sigma_x}\right)+\frac{1}{\sigma_x}\log_2(x-1)=1 .
\end{equation}
For an integer $k\ge2$, we write $\sigma_k\triangleq\sigma_x\big|_{x=k}$, and $\varepsilon_k\triangleq\varepsilon_x\big|_{x=k}$.

Thus, $\varepsilon_k\in(0,\frac12]$ is the unique solution of Equation~\eqref{eq:H=1} with $x=k$, and $\sigma_k=\frac{1}{\varepsilon_k}\ge2$. 
The following lemma records elementary properties of $\sigma_x$ and $\varepsilon_x$ that are used below.

\begin{lemma}\label{lemma:sigma}
For every real $x\ge2$, let $\sigma_x\ge2$ and $\varepsilon_x=\frac{1}{\sigma_x}$ be defined by Equation~\eqref{eq:sigma}, and for integer $k\ge2$, let $\sigma_k$ and $\varepsilon_k$ be their restrictions as above. Then
\begin{enumerate}
\item[$(1)$] $\sigma_2=2$;
\item[$(2)$] $\sigma_x\ \ge\ 2+\log_2(x-1)$ for all real $x\ge2$;
\item[$(3)$] $\displaystyle\lim_{k\to\infty}\varepsilon_k=0$.
\end{enumerate}
\end{lemma}
\begin{proof}
\begin{enumerate}
\item[$(1)$]For $x=2$, Equation~\eqref{eq:sigma} reads $h(1/\sigma_2)=1$. Since $h$ is strictly increasing on
	$[0,\frac12]$ and $h(\frac12)=1$, we obtain $1/\sigma_2=\frac12$; that is, $\sigma_2=2$.
\item[$(2)$] 
	Since $h$ is concave on $[0,\frac12]$ with $h(0)=0$ and $h(\frac12)=1$, its graph lies above
	the chord joining $(0,0)$ and $(\frac12,1)$; that is, above the line $y=2t$. 
	Therefore, we obtain $h(t)\ge2t$ for $t\in[0,\frac12]$.
	Because $\varepsilon_x\in(0,\frac12]$, it follow that
	\[
	1=h(\varepsilon_x)+\varepsilon_x\log_2(x-1)
	\ \ge\ 2\varepsilon_x+\varepsilon_x\log_2(x-1)
	=\varepsilon_x\big(2+\log_2(x-1)\big),
	\]
and hence $2+\log_2(x-1)\le \frac{1}{\varepsilon_x} =\sigma_x$ for all real $x\ge2$.
\item[$(3)$] By $(2)$, for every integer $k\ge2$, 
	\[
	0\ <\ \varepsilon_k\ \le\ \frac{1}{2+\log_2(k-1)}.
	\]
Since $\log_2(k-1)\to\infty$ as $k\to\infty$, the upper bound tends to $0$. Hence, $\lim_{k\to\infty}\varepsilon_k=0$ by the squeeze theorem.
\end{enumerate}
\end{proof}

At the end of this subsection,
we prove a lemma concerning the one-point plus uniform-tail distributions.
Before stating the lemma, we introduce the following notation.
\begin{equation} \label{notation1}
   \begin{aligned}
   R_k(\varepsilon) & \triangleq \dfrac{A_{P_{\varepsilon,k}}(L)}{\max\{1,H(P_{\varepsilon,k})\}},  \\
   \overline{B}  & \triangleq  \sup_{\substack{2\leq k\in\mathbb{N}\\ 0<\varepsilon\leq(k-1)/k}} R_k(\varepsilon)=\sup_{P\in\mathcal{P}(\varepsilon,k)}\dfrac{A_{P}(L)}{\max\{1,H(P)\}}. \\                                         \end{aligned}
\end{equation}
\begin{lemma} \label{lemma1}
For any prefix code $\mathcal{C}$, the following hold:
\begin{enumerate}
    \item[$(1)$] $\overline{B}\geq 1+L(1),$
    \item[$(2)$] $\lim_{\varepsilon\rightarrow 0^+}R_k(\varepsilon)=L(1)$ for any fixed integer $k\geq2$,
    \item[$(3)$] $a_k\leq \overline{B}\log_{2}k$ for any integer $k\geq2$.
\end{enumerate}
\end{lemma}
\begin{proof}
\begin{enumerate}
    \item[$(1)$] Since $g(x)=2^{-x}$ is a convex function, Jensen's inequality yields
    \[
     2^{-\widetilde{a}_k}=2^{-\frac{\sum_{n=2}^{k}L(n)}{k-1}}\leq \frac{1}{k-1}\sum_{n=2}^{k}2^{-L(n)}\leq\frac{1}{k-1},
    \]
    where the last inequality follows from the Kraft inequality~\cite{kraft}.
    This further yields $\widetilde{a}_k\geq \log_{2}(k-1)$ for all $2\leq k\in\mathbb{N}$.
    By the definition of $\overline{B}$, we have
  \begin{equation*}
  \begin{aligned}
  \overline{B}  & \geq R_k(\varepsilon_k)=A_{P_{\varepsilon_k,k}}(L)    \\
                &  = (1-\varepsilon_k)L(1) + \varepsilon_k \widetilde{a}_k     \\
                & \geq (1-\varepsilon_k)L(1) + \varepsilon_k \log_{2}(k-1)  \\
                & = (1-\varepsilon_k)L(1) +1-h(\varepsilon_k)   \\
  \end{aligned}
  \end{equation*}
  for all $2\leq k\in\mathbb{N}$. Letting $k\rightarrow\infty$,
  we obtain from $\lim_{k\rightarrow\infty}\varepsilon_k=0$ that $h(\varepsilon_k)\rightarrow 0$, and hence
  \[
    \overline{B}\geq \lim_{k\rightarrow\infty}\big[(1-\varepsilon_k)L(1) +1-h(\varepsilon_k)\big]= 1+L(1).
  \]
    \item[$(2)$] For any fixed integer $k\geq2$, if $\varepsilon\in(0,\varepsilon_k]$, then $H(P_{\varepsilon,k})\leq1$, and hence
    $\max\{1,H(P_{\varepsilon,k})\}=1$. We obtain
    \begin{equation*}
  \begin{aligned}
  \lim_{\varepsilon\rightarrow 0^+}R_k(\varepsilon)  &=  \lim_{\varepsilon\rightarrow 0^+}A_{P_{\varepsilon,k}}(L)   \\
                &  =  \lim_{\varepsilon\rightarrow 0^+} \big[ (1-\varepsilon)L(1) + \varepsilon \widetilde{a}_k \big]   \\
                & = L(1)  \\
  \end{aligned}
  \end{equation*}
  for any fixed integer $k\geq2$.
    \item[$(3)$] Since $U_k=P_{\frac{k-1}{k},k}\in \mathcal{P}(\varepsilon,k)$, it follows that
    \[
     \frac{a_k}{\log_2{k}}=\dfrac{A_{U_k}(L)}{\max\{1,H(U_k)\}}\leq \overline{B}
    \]
    for all $2\leq k\in\mathbb{N}$, and hence $a_k\leq \overline{B}\log_2 k$ holds for every integer $k\geq2$.
\end{enumerate}
\end{proof}
Naturally, we obtain the following corollary, which is Theorem~\ref{thm1}.
\begin{corollary} \label{cor1}
Let $\mathcal{C}$ be an arbitrary UCI.
Then its minimum expansion factor satisfies $C_{\mathcal{C}}^{*}\geq 1+L_{\mathcal{C}}(1)$, which further implies that
$C_{\mathcal{C}}^{*}\geq2$ holds for every UCI $\mathcal{C}$.
\end{corollary}
\begin{proof}
From the definitions of $C_{\mathcal{C}}^{*}$ and $\overline{B}$, and by Lemma~\ref{lemma1}$(2)$, we have
\[
  C_{\mathcal{C}}^{*}=\sup_{P\in\mathcal{P}_\mathbb{N}}\dfrac{A_{P}(L_\mathcal{C})}{\max\{1,H(P)\}}\geq \sup_{P\in\mathcal{P}(\varepsilon,k)}\dfrac{A_{P}(L_\mathcal{C})}{\max\{1,H(P)\}}=\overline{B}\geq 1+L_{\mathcal{C}}(1)\geq2.
\]
\end{proof}
\subsection{Least favorable distributions of \(\mathcal{P}_\mathbb{N}\)}
We prove an important result: for any prefix code $\mathcal{C}$, every least favorable distribution from the nonincreasing family $\mathcal{P}_\mathbb{N}$ must be a one-point plus uniform-tail distribution. 
Here, $P\in\mathcal{P}_\mathbb{N}$ is called least favorable for $\mathcal{C}$ if it attains the worst-case ratio; that is, $A_P(L_{\mathcal{C}})/\max\{1,H(P)\}=C_{\mathcal{C}}^{*}$, where $C_{\mathcal{C}}^{*}$ is the supremum of this ratio over all $P\in\mathcal{P}_\mathbb{N}$.
To establish this result, we first consider a function $\Phi$ and establish an auxiliary lemma.

Define the function $\Phi(y)$ as follows.
\begin{equation}  \label{function}
\Phi(y)\triangleq y\left(1-h\left(\frac{1}{y}\right)\right),\qquad y\ge 2.
\end{equation}
Taking the derivative yields
\[
\Phi'(y)=1+\log_2\left(1-\frac1y\right)> 0
\]
for all $y>2$.
In fact, \(\Phi'(2)=0\) does not affect the fact that \(\Phi\) is strictly increasing on \([2,\infty)\). Moreover,
\[
\Phi''(y)=\frac{1}{y(y-1)\ln 2}>0
\]
for all $y\geq2$. Thus, \(\Phi\) is strictly convex on \([2,\infty)\).
We also have $\Phi(2)=0$ and $\lim_{y\rightarrow\infty}\Phi(y)=\infty $.
Thus $\Phi:[2,\infty)\longrightarrow[0,\infty)$ is a strictly increasing bijection.
Therefore, we can define the inverse function
\begin{equation} \label{function Y}
Y\triangleq \Phi^{-1}:[0,\infty)\longrightarrow[2,\infty).
\end{equation}
Because the inverse of a strictly increasing function is itself strictly increasing,
it follows that \(Y\) is strictly increasing on \([0, \infty)\).
For \(u>0\), we have \(Y(u)>Y(0)=2\), so the derivative property of inverse functions gives
\begin{equation} \label{eq6}
Y'(u)=\frac{1}{\Phi'(Y(u))}=\dfrac{1}{1+\log_{2}(1-\frac{1}{Y(u)})}.
\end{equation}
Differentiating again yields
\[
Y''(u)
=-\frac{\Phi''(Y(u))}{\Phi'(Y(u))^3}<0
\]
for all $u>0$. Hence \(Y'\) is strictly decreasing on \((0,\infty)\).
Next, we prove \(Y\) is strictly concave on \([0,\infty)\).\footnote{Note that to prove that \(Y\) is strictly concave on \((0,\infty)\), it suffices to show that \(Y''(u) < 0\) for all $u>0$.}
Take $0\le u_1<u_2$ and $0<\lambda<1$. Let $y_1\triangleq Y(u_1)$ and $y_2\triangleq Y(u_2)$.
Since \(\Phi\) is strictly convex on \([2,\infty)\), we have
\[
\Phi\bigl((1-\lambda)y_1+\lambda y_2\bigr)
<
(1-\lambda)\Phi(y_1)+\lambda\Phi(y_2)=(1-\lambda)u_1+\lambda u_2.
\]
Applying the strictly increasing inverse \(Y\) to both sides gives $(1-\lambda)y_1+\lambda y_2 < Y\bigl((1-\lambda)u_1+\lambda u_2\bigr)$.
Equivalently,
\[
(1-\lambda)Y(u_1)+\lambda Y(u_2)
<
Y\bigl((1-\lambda)u_1+\lambda u_2\bigr).
\]
This is exactly the definition of strict concavity. Thus \(Y\) is strictly concave on \([0,\infty)\).
We summarize the relevant results on the functions $\Phi$ and \(Y\) as follows:
\begin{enumerate}
    \item $\Phi$ is strictly increasing on \([2,\infty)\);
    \item \(\Phi\) is strictly convex on \([2,\infty)\);
    \item \(Y\) is strictly increasing on \([0,\infty)\);
    \item \(Y'\) is strictly decreasing on \((0,\infty)\);
    \item \(Y\) is strictly concave on \([0,\infty)\).
\end{enumerate}
We now present the auxiliary lemma.
\begin{lemma}\label{lemma2}
Let \(P \in \mathcal{P}_\mathbb{N}\) be a probability distribution and define
$w_k \triangleq k\big(P(k)-P(k+1)\big)$ for all $k\in\mathbb{N}$.
Then the following statements hold.
\begin{enumerate}
    \item[$(1)$] $W\triangleq(w_1,w_2,\ldots)$ is a probability distribution, i.e., $w_k\ge0$ for all $k\in\mathbb{N}$ and $\sum_{k=1}^{\infty}w_k=1$.
\item[$(2)$] For every \(n\ge1\),
\[
P(n)=\sum_{k=n}^{\infty}\frac{w_k}{k},
\]
and therefore $P=\sum_{k=1}^{\infty}w_kU_k$.\footnote{The expression $\sum_{k\ge1}w_kU_k$ denotes a linear combination of probability distributions. This combination can be written as $\Big(\sum_{k\ge1}w_kU_k(1),\ \sum_{k\ge1}w_kU_k(2),\ \dots\Big)$. Consequently, $P=\sum_{k\ge1}w_kU_k$ is equivalent to
$P(n)=\sum_{k\ge1}w_kU_k(n)$ for every $n\in\mathbb N$.}
The average codeword length is $A_P(L)=\sum_{k=1}^{\infty}w_k a_k$.
    \item[$(3)$]  The entropy satisfies
    \[
H(P)=\sum_{k=1}^{\infty}w_k\log_2 k+\sum_{k=1}^{\infty}w_kD(U_k\|P),
\]
where $D(U_k\|P)$ denotes the relative entropy of $U_k$ with respect to $P$.
In particular, $$H(P)\ge\sum_{k=1}^{\infty}w_k\log_2 k.$$
\end{enumerate}
\end{lemma}
\begin{proof}
\begin{enumerate}
\item[$(1)$]Since \(P\) is nonincreasing, $w_k=k\big(P(k)-P(k+1)\big)\ge0$ for all $k\in\mathbb{N}$.
For every \(n\ge1\),
\[
\begin{aligned}
\sum_{k=1}^n w_k &= \sum_{k=1}^n kP(k)-\sum_{k=1}^n kP(k+1) \\
                 &= P(1)-nP(n+1)+\sum_{k=2}^n P(k)[k-(k-1)] \\
                 &= \sum_{k=1}^n P(k)-nP(n+1).
\end{aligned}
\]
Since $\lim_{n\to\infty}\sum_{k=1}^n P(k)=\sum_{k=1}^\infty P(k)=1$, we obtain
\[
\lim_{m\to\infty}\sum_{k=m}^\infty P(k)=\lim_{m\rightarrow\infty}\left[\sum_{k=1}^\infty P(k)-\sum_{k=1}^{m-1} P(k)\right]=1-\lim_{m\rightarrow\infty}\sum_{k=1}^{m-1}P(k)=0
\]
Since \(P\) is nonincreasing, for every \(n\in\mathbb{N}\) we have
\begin{equation*}
 0\leq nP(n+1)\leq 2\sum_{k=\lceil\frac{n}{2}\rceil}^{n+1}P(k)  \leq  2\sum_{k=\lceil\frac{n}{2}\rceil}^\infty P(k),
\end{equation*}
and letting $n\to\infty$ yields $\lim_{n\to\infty}nP(n+1)=0$. Therefore
\[
\sum_{k=1}^\infty w_k=\lim_{n\to\infty}\sum_{k=1}^n w_k=\lim_{n\to\infty}\left[\sum_{k=1}^n P(k)-nP(n+1)\right] =1.
\]
\item[$(2)$]
For every $n\in\mathbb{N}$, we obtain
\[
\begin{aligned}
\sum_{k=n}^\infty \frac{w_k}{k}& =\sum_{k=n}^\infty \Big(P(k)-P(k+1)\Big)  \\
&= \lim_{m\to\infty}\Big(P(n)-P(m)\Big)  \\
&=P(n).
\end{aligned}
\]
Furthermore, we have
\[
  \sum_{k=1}^{\infty}w_kU_k(n)=\sum_{k=1}^{n-1}w_k\cdot 0+\sum_{k=n}^{\infty}w_k\cdot\frac{1}{k}=P(n)
\]
for all $n\in\mathbb{N}$. Therefore, $P=\sum_{k=1}^{\infty}w_kU_k$.
The average codeword length is
\[
\begin{aligned}
A_P(L) &= \sum_{n=1}^\infty L(n)P(n)= \sum_{n=1}^\infty L(n) \sum_{k=n}^{\infty}\frac{w_k}{k}  \\
       &= \sum_{k=1}^\infty \frac{w_k}{k} \sum_{n=1}^k L(n)= \sum_{k=1}^\infty   w_k a_k.\\
\end{aligned}
\]
\item[$(3)$]
Define an auxiliary random variable \(K\) with $K\sim W$, i.e.,
\[
\Pr(K=k)=w_k, \qquad k\in\mathbb{N}.
\]
Define a random variable \(X\) such that the joint distribution of \(X\) and \(K\) satisfies
\[
\Pr(X=n,K=k)=w_kU_k(n), \qquad n,k\in\mathbb{N}.
\]
Then, the distribution of \(X\) is
\[
\begin{aligned}
\Pr(X=n) &= \sum_{k=1}^\infty \Pr(X=n,K=k)  \\
&=\sum_{k=1}^\infty w_kU_k(n)\\
&=P(n)
\end{aligned}
\]
for all $n\in\mathbb{N}$. Hence $X\sim P$.
The conditional probability of \(X=n\) given \(K=k\) is
\[
\Pr(X=n\mid K=k)=\dfrac{\Pr(X=n,K=k)}{\Pr(K=k)}=U_k(n)  \\
\]
for all $n,k\in\mathbb{N}$.
Thus, the conditional entropy of \(X\) given \(K\) is
\[
\begin{aligned}
H(X\mid K) &= \sum_{k=1}^\infty \Pr(K=k) H(X\mid K=k)\\
           &= \sum_{k=1}^\infty w_kH(U_k)\\
           &= \sum_{k=1}^\infty w_k\log_2 k.
\end{aligned}
\]
Next, the mutual information
\[
\begin{aligned}
I(K;X) &= \sum_{k=1}^\infty \sum_{n=1}^\infty w_kU_k(n)\log_2\frac{w_kU_k(n)}{w_kP(n)}\\
       &= \sum_{k=1}^\infty w_k \sum_{n=1}^\infty U_k(n) \log_2\frac{U_k(n)}{P(n)}\\
       &= \sum_{k=1}^\infty w_kD(U_k\|P).
\end{aligned}
\]
By Theorem~\ref{thm2}, it follows that $H(X)=H(X\mid K)+I(K;X)$. Since \(X\sim P\), we have \(H(X)=H(P)\). Therefore
\[
\begin{aligned}
H(P)
&=H(X)\\
&=H(X\mid K)+I(K;X)\\
&=\sum_{k=1}^{\infty}w_k\log_2 k+\sum_{k=1}^{\infty}w_kD(U_k\|P).
\end{aligned}
\]
Since the relative entropy $D(U_k\|P)\ge0$, we conclude that
\[
H(P)\ge\sum_{k=1}^\infty w_k\log_2 k.
\]
\end{enumerate}
\end{proof}
When \(Q\) is a nonincreasing probability distribution on the alphabet \(\{2,3, \ldots\}\), a similar result holds as follows.
\begin{lemma}~\label{lemma3}
Let $Q$ be a nonincreasing probability distribution on the alphabet $\{2,3,\dots\}$ with $H(Q)<\infty$.
Define $\widetilde{w}_k \triangleq (k-1)\big(Q(k)-Q(k+1)\big)$ for all $2\leq k\in\mathbb{N}$.
Then the following statements hold.
\begin{enumerate}
	\item[$(1)$] $\widetilde{W}\triangleq(\widetilde{w}_2,\widetilde{w}_3,\ldots)$ is a probability distribution, i.e., $\widetilde{w}_k\ge0$ for all $2\leq k\in\mathbb{N}$ and $\sum_{k=2}^{\infty}\widetilde{w}_k=1$.
	\item[$(2)$] The probability distribution $Q=\sum_{k=2}^{\infty}\widetilde{w}_k\widetilde{U}_k$. The average codeword length is $A_Q(L)=\sum_{k=2}^{\infty}\widetilde{w}_k \widetilde{a}_k$.
	\item[$(3)$]  The entropy satisfies
	\[
	H(Q)=\sum_{k=2}^{\infty}\widetilde{w}_k\log_2 (k-1)+\sum_{k=2}^{\infty}\widetilde{w}_kD(\widetilde{U}_k\|Q),
	\]
	where $D(\widetilde{U}_k\|Q)$ denotes the relative entropy of $\widetilde{U}_k$ with respect to $Q$.
	In particular, $$H(Q)\ge\sum_{k=2}^{\infty}\widetilde{w}_k\log_2 (k-1).$$
\end{enumerate}
\end{lemma}
The proof of Lemma~\ref{lemma3} is identical to that of Lemma~\ref{lemma2}, except that the alphabet is \(\{2,3,\ldots\}\) rather than $\mathbb{N}$, so that all sums start at $2$; we therefore omit it.
Lemmas~\ref{lemma2} and~\ref{lemma3} show that every non-increasing probability distribution on the alphabet can be written as a linear combination of the corresponding uniform distributions, where the linear coefficients precisely form a probability distribution.
At the end of this subsection, we present the main theorem of this section.
\begin{theorem} \label{thm3}
Let $\mathcal{C}$ be any prefix code. Then
\begin{equation}\label{eq3}
 \sup_{P\in\mathcal{P}_\mathbb{N}}\dfrac{A_{P}(L_\mathcal{C})}{\max\{1,H(P)\}} = \sup_{P\in\mathcal{P}(\varepsilon,k)}\dfrac{A_{P}(L_\mathcal{C})}{\max\{1,H(P)\}};
\end{equation}
that is, $C_{\mathcal{C}}^{*}=\overline{B}$.
\end{theorem}
\begin{proof}
Since $\mathcal{P}(\varepsilon,k) \subseteq \mathcal{P}_\mathbb{N}$, it follows that $C_{\mathcal{C}}^{*}\geq \overline{B}$.
Thus, we only need to prove that $C_{\mathcal{C}}^{*}\leq \overline{B}$.

If $\overline{B}=+\infty$, then $C_{\mathcal{C}}^{*}\geq \overline{B}=+\infty$, which implies that Equation~\eqref{eq3} holds.
Therefore, it remains to prove that if $\overline{B}<+\infty$, then
\begin{equation}  \label{eq4}
    \dfrac{A_{P}(L_\mathcal{C})}{\max\{1,H(P)\}}\leq \overline{B}
\end{equation}
for all $P\in\mathcal{P}_\mathbb{N}$. We consider two cases depending on the value of \(P(1)\).
\begin{enumerate}
\item Case $P(1)\leq 2^{-\frac{L(1)}{\overline{B}}}$:
By Lemma~\ref{lemma2}, 
there exists \(\{w_k\}_{k \ge 1}\) with \(w_k \ge 0\) and \(\sum_{k=1}^{\infty} w_k = 1\) such that \(P =\sum_{k=1}^\infty w_k U_k\) and $H(P)=\sum_{k=1}^{\infty}w_k\log_2 k+\sum_{k=1}^{\infty}w_kD(U_k\|P)$.
Because $P(1)\leq 2^{-\frac{L(1)}{\overline{B}}}$ and $w_kD(U_k\|P)\geq 0$ for all $k\in\mathbb{N}$, it follows that
\[
 \begin{aligned}
H(P)-\sum_{k=1}^{\infty}w_k\log_2 k & = \sum_{k=1}^{\infty}w_kD(U_k\|P)\\
                                    & \geq w_1 D(U_1\|P)\\
                                    & =-w_1 \log_2 P(1)  \\
                                    & \geq \frac{w_1L(1)}{\overline{B}}.
\end{aligned}
\]
Furthermore, by Lemmas~\ref{lemma1} and~\ref{lemma2}, we obtain
\[
 \begin{aligned}
 A_P(L) & =\sum_{k=1}^{\infty}w_k a_k  \\
        & \leq  w_1a_1+\overline{B}\sum_{k=2}^\infty w_k \log_2 k          \\
        & = w_1 L(1)+\overline{B}\sum_{k=2}^\infty w_k \log_2 k    \\
        & = \overline{B}\left(\frac{w_1 L(1)}{\overline{B}}+\sum_{k=2}^\infty w_k \log_2 k  \right)  \\
        & \leq \overline{B}H(P).
 \end{aligned}
\]
Therefore, we have
\[
 \dfrac{A_{P}(L)}{\max\{1,H(P)\}}\leq \dfrac{A_{P}(L)}{H(P)} \leq   \overline{B}.
\]
\item Case $P(1)> 2^{-\frac{L(1)}{\overline{B}}}$:
By the definition of \(\overline{B}\) and \(H(P_{\varepsilon_k,k}) = 1\), it follows that
\[
  A_{P_{\varepsilon_k,k}}(L)=(1-\varepsilon_k)L(1)+\varepsilon_k \widetilde{a}_k\leq \overline{B}
\]
then
\[
 \widetilde{a}_k \leq L(1)+\left(\overline{B}-L(1)\right)\sigma_k
\]
for all $k\geq2$.
Let \(Y\) be the function defined in Equation~\eqref{function Y}. Then \(\sigma_k=Y\bigl(\log_2(k-1)\bigr)\).

Let $q\triangleq 1-P(1)$ and $Q(n)\triangleq \frac{P(n)}q$ for all $2\leq n\in\mathbb{N}$.
It is clear that $Q$ is a nonincreasing probability distribution on the alphabet $\{2,3,\dots\}$ with $H(Q)<\infty$.
By Lemma~\ref{lemma3}, there exists a probability distribution \(\widetilde{W}\triangleq(\widetilde{w}_2,\widetilde{w}_3,\ldots)\) such that
$Q=\sum_{k=2}^{\infty}\widetilde{w}_k\widetilde{U}_k$, $A_Q(L)=\sum_{k=2}^{\infty}\widetilde{w}_k \widetilde{a}_k$ and $H(Q)\ge\sum_{k=2}^{\infty}\widetilde{w}_k\log_2 (k-1)$.
Therefore, we obtain
\[
  \begin{aligned}
  A_P(L)& =P(1)L(1)+[1-P(1)]\sum_{n=2}^\infty Q(n)L(n)    \\
        & =P(1)L(1)+q \sum_{n=2}^\infty \widetilde{w}_k \widetilde{a}_k  \\
        & \leq P(1)L(1)+ q \sum_{n=2}^\infty \widetilde{w}_k\left[L(1)+\left(\overline{B}-L(1)\right)\sigma_k\right]  \\
        &= L(1)+q\left(\overline{B}-L(1)\right)\sum_{n=2}^\infty \widetilde{w}_k Y\bigl(\log_2(k-1)\bigr).
   \end{aligned}
\]
Because \(Y\) is concave, Jensen's inequality yields
\[
\sum_{n=2}^\infty \widetilde{w}_k Y\bigl(\log_2(k-1)\bigr) \le Y\left(\sum_{n=2}^\infty  \widetilde{w}_k\log_2(k-1)\right).
\]
Since \(Y\) is increasing and $H(Q)\ge\sum_{k=2}^{\infty}\widetilde{w}_k\log_2 (k-1)$, we have
\[
Y\left(\sum_{k=2}^{\infty}\widetilde{w}_k\log_2 (k-1)\right) \le Y(H(Q)).
\]
Therefore, we obtain
\[
  \begin{aligned}
  A_P(L)& \leq  L(1)+q\left(\overline{B}-L(1)\right)\sum_{n=2}^\infty \widetilde{w}_k Y\bigl(\log_2(k-1)\bigr)    \\
        & \leq L(1)+q\left(\overline{B}-L(1)\right)Y\left(\sum_{n=2}^\infty  \widetilde{w}_k\log_2(k-1)\right)\\
        & \leq L(1)+q\left(\overline{B}-L(1)\right)Y(H(Q))  \\
   \end{aligned}
\]
Let $u\triangleq H(Q)$, then
\[
 H(P)=(1-P(1))H(Q)+h(P(1))=qu+h(q).
\]
To prove Equation~\eqref{eq4}, it suffices to prove that
\begin{equation} \label{eq5}
  L(1)+q\left(\overline{B}-L(1)\right)Y(u)\leq \overline{B}\max\{1,qu+h(q)\}.
\end{equation}
We prove Equation~\eqref{eq5} by considering two cases depending on the value of \(H(P)\).
\begin{enumerate}
\item Case $H(P)=qu+h(q)\leq1$: Let $u_0 \triangleq \frac{1-h(q)}{q}$, then $u\leq u_0$.
Let \(\Phi\) be the function defined in Equation~\eqref{function}.
We have $\Phi(\frac{1}{q})=\frac{1}{q}[1-h(q)]=u_0$.
\footnote{By Lemma~\ref{lemma1}, $\overline{B}\geq 1+L(1)$. Since $P(1)>2^{-\frac{L(1)}{\overline{B}}}$, we obtain $P(1)>2^{-\frac{L(1)}{\overline{B}}}>2^{-\frac{L(1)}{L(1)}}=\frac{1}{2}$. Therefore, \(\frac{1}{q}>2\), which lies in the domain of \(\Phi\).}
Since \(Y\) is increasing,
\[
  Y(u)\leq Y(u_0)=Y\Big(\Phi(\frac{1}{q})\Big)=\frac{1}{q},
\]
and hence
\[
  L(1)+q\left(\overline{B}-L(1)\right)Y(u)\leq  L(1)+q\left(\overline{B}-L(1)\right)\frac{1}{q}= \overline{B}\max\{1,qu+h(q)\}.
\]
\item Case $H(P)=qu+h(q)>1$: In this case, $u>u_0$.
Let $$F(u)\triangleq \overline{B}\big[qu+h(q)\big]-L(1)-q\left(\overline{B}-L(1)\right)Y(u),$$
\end{enumerate}
then $F(u_0)=\overline{B}-L(1)-q\left(\overline{B}-L(1)\right)\frac{1}{q}=0$.
From Equation~\eqref{eq6}, it follows that $Y'(u)=\big(1+\log_{2}(1-\frac{1}{Y(u)})\big)^{-1}$.
Since \(Y'\) is decreasing and $P(1)=1-q>2^{-\frac{L(1)}{\overline{B}}}$, we obtain
\[
   \begin{aligned}
      Y'(u) & \leq Y'(u_0)=\frac{1}{1+\log_{2}(1-q)}   \\
            & < \frac{1}{1-\frac{L(1)}{\overline{B}}} = \frac{\overline{B}}{\overline{B}-L(1)}
   \end{aligned}
\]
for all $u\geq u_0$.
Therefore, taking the derivative of \(F(u)\) yields
\[
  \begin{aligned}
   F'(u) & = \overline{B}q-q\left(\overline{B}-L(1)\right)Y'(u)   \\
         & > \overline{B}q-q\left(\overline{B}-L(1)\right)\frac{\overline{B}}{\overline{B}-L(1)}  \\
         & =0,
   \end{aligned}
\]
and hence $F(u)$ is increasing on $[u_0,\infty)$.
Thus $F(u)>F(u_0)=0$, which implies that Equation~\eqref{eq5} holds.
\end{enumerate}
This completes the proof.
\end{proof}
Theorem~\ref{thm3} shows that, regardless of how the prefix code \(\mathcal{C}\) is constructed, at least one of the one-point plus uniform-tail distributions attains the supremum of the ratio $\frac{A_{P}(L)}{\max\{1,H(P)\}}$.
	
\subsection{Least favorable distributions of \(\mathcal{P}(\varepsilon,k)\)}
In this subsection, we prove that for any prefix code $\mathcal{C}$, the least favorable distributions chosen from the family \(\mathcal{P}(\varepsilon,k)\) must belong to $\{P_{\varepsilon_k,k},U_k \mid 2\leq k\in\mathbb{N}\}$.
We now prove an important lemma.
\begin{lemma}  \label{lemma5}
Let $\mathcal{C}$ be any prefix code. Then
\begin{equation} \label{eq7}
 \sup_{0<\varepsilon\leq(k-1)/k}\frac{A_{P_{\varepsilon,k}}(L)}{\max\{1,H(P_{\varepsilon,k})\}}=\max \left\{(1-\varepsilon_k)L(1)+\varepsilon_k\widetilde{a}_k, \frac{a_k}{\log_2 k} \right\}
\end{equation}
for all $2\leq k\in\mathbb{N}$.
\end{lemma}
\begin{proof}
Let
\[
   \begin{aligned}
   \rho_B(\varepsilon) & \triangleq \frac{B\max\{1,H(P_{\varepsilon,k})\}-(1-\varepsilon)L(1)}{\varepsilon}.  \\
   \end{aligned}
\]
The proof has three parts. We first prove that
\begin{equation} \label{eq8}
 \sup_{0<\varepsilon\leq(k-1)/k} R_k(\varepsilon)\leq B\Longleftrightarrow \widetilde{a}_k \leq \inf_{0<\varepsilon\leq(k-1)/k}\rho_B(\varepsilon)
\end{equation}
holds for any constant \(B\).
We then prove that equation
\begin{equation} \label{eq9}
\inf_{0<\varepsilon\le(k-1)/k}\rho_B(\varepsilon)= \min \left\{\rho_B(\varepsilon_k),\rho_B\left(\frac{k-1}{k}\right)\right\}.
\end{equation}
    holds for any constant \(B\) satisfying $\sup_{0<\varepsilon\leq(k-1)/k} R_k(\varepsilon)\leq B$.
Finally, we prove that Equation~\eqref{eq7} holds.
\begin{enumerate}
\item For fixed \(B\) and \(\varepsilon\), we have
\[
  \begin{aligned}
  R_k(\varepsilon)\leq B & \Longleftrightarrow A_{P_{\varepsilon,k}}(L)=(1-\varepsilon)L(1)+\varepsilon\widetilde a_k\leq B\max\{1,H(P_{\varepsilon,k})\}  \\ & \Longleftrightarrow \widetilde a_k \leq \rho_B(\varepsilon).
  \end{aligned}
\]
Therefore, we obtain
\[
  \forall \varepsilon\in\left(0,\frac{k-1}{k}\right],R_k(\varepsilon)\leq B \Longleftrightarrow  \forall \varepsilon\in\left(0,\frac{k-1}{k}\right],\widetilde a_k \leq \rho_B(\varepsilon)
\]
for all constant \(B\); that is, Equation~\eqref{eq8} holds.
\item Since $f(x)=h(x)+x\log_{2}(k-1)$ is strictly increasing on $(0,\frac{k-1}{k}]$, and $\varepsilon_k$ is the unique solution of \(f(x)=1\), it follows that \(f(\varepsilon)\leq 1\) for \(\varepsilon\in(0,\varepsilon_k]\), and \(f(\varepsilon)>1\) for \(x\in(\varepsilon_k,\frac{k-1}{k}]\).
\begin{enumerate}
\item Case \(\varepsilon\in(0,\varepsilon_k]\): In this case, $H(P_{\varepsilon,k})=h(\varepsilon)+\varepsilon\log_{2}(k-1)\leq 1$,
 and hence
 \[
  \rho_B(\varepsilon)=  \frac{B-(1-\varepsilon)L(1)}{\varepsilon}=L(1)+\frac{B-L(1)}{\varepsilon}.
 \]
 Also, since
 \[
  B\geq  R_k(\varepsilon)= A_{P_{\varepsilon,k}}(L)= \sum_{n=1}^k P_{\varepsilon,k}(n)L(n)\geq \sum_{n=1}^k P_{\varepsilon,k}(n)L(1)=L(1),
 \]
 it follows that
 \[
  \rho_B(\varepsilon)=L(1)+\frac{B-L(1)}{\varepsilon}\geq L(1)+\frac{B-L(1)}{\varepsilon_k}=\rho_B(\varepsilon_k)
 \]
 for all $\varepsilon\in(0,\varepsilon_k]$. Therefore, we obtain $\inf_{0<\varepsilon\le\varepsilon_k}\rho_B(\varepsilon)=\rho_B(\varepsilon_k)$.
\item Case \(\varepsilon\in(\varepsilon_k,\frac{k-1}{k}]\): In this case,\footnote{When $k=2$, we have $\varepsilon_k=\frac{k-1}{k}=\frac{1}{2}$. Therefore, this case exists for \(k\ge 3\).} we obtain
\[
 \rho_B(\varepsilon)=  \frac{B\big(h(\varepsilon)+\varepsilon\log_{2}(k-1)\big)-(1-\varepsilon)L(1)}{\varepsilon}=B\log_{2}(k-1)+L(1)+
\frac{Bh(\varepsilon)-L(1)}{\varepsilon}.
\]
Taking the derivative yields
\[
\rho_B'(\varepsilon)=\frac{B\log_2(1-\varepsilon)+L(1)}{\varepsilon^2}.
\]
The numerator $B\log_2(1-\varepsilon)+L(1)$ is strictly decreasing in \(\varepsilon\) and has the unique zero
\[
\varepsilon_B \triangleq 1-2^{-L(1)/B}\leq 1-2^{-1}=\frac{1}{2}.
\]
Hence \(\rho_B\) first increases and then decreases on $(\varepsilon_k,\frac{k-1}{k}]$. Thus, we obtain
\[
\inf_{\varepsilon_k\le\varepsilon\le(k-1)/k}\rho_B(\varepsilon)=\min\left\{\rho_B(\varepsilon_k),\rho_B\left(\frac{k-1}{k}\right)\right\}.
\]
\end{enumerate}
Combining the two cases above, we conclude that Equation~\eqref{eq9} holds.
\item At \(\varepsilon=\varepsilon_k\), the entropy is \(1\), and therefore $\rho_B(\varepsilon_k)=\frac{B-(1-\varepsilon_k)L(1)}{\varepsilon_k}$. Thus
\[
\widetilde a_k\le\rho_B(\varepsilon_k)
\quad\Longleftrightarrow\quad (1-\varepsilon_k)L(1)+\varepsilon_k\widetilde a_k \le B.
\]

At \(\varepsilon=(k-1)/k\), the distribution becomes the uniform distribution \(U_k\), and we have $A_{P_{\varepsilon,k}}(L)=a_k$ and
$H(P_{\varepsilon,k})=\log_2 k$. A direct computation gives
\[
\widetilde a_k
\le
\rho_B\left(\frac{k-1}{k}\right)
\quad\Longleftrightarrow\quad
\frac{a_k}{\log_2k}
\le B.
\]
Hence, it follows that
\[
 \begin{aligned}
 \sup_{0<\varepsilon\leq(k-1)/k} R_k(\varepsilon)\leq B & \Longleftrightarrow \widetilde{a}_k \leq \inf_{0<\varepsilon\leq(k-1)/k}\rho_B(\varepsilon)   \\
                  & \Longleftrightarrow  \widetilde{a}_k \leq \min\left\{\rho_B(\varepsilon_k), \ \rho_B\left(\frac{k-1}{k}\right)\right\}\\
                  & \Longleftrightarrow  \max\left\{(1-\varepsilon_k)L(1)+\varepsilon_k\widetilde a_k ,\ \frac{a_k}{\log_2k}\right\}\leq B. \\
  \end{aligned}
\]
Therefore, Equation~\eqref{eq7} holds for all \(2\leq k\in\mathbb{N}\).
\end{enumerate}
This completes the proof.
\end{proof}
The following corollary is an immediate consequence of Lemma~\ref{lemma5}.
\begin{corollary} \label{cor2}
Let $\mathcal{C}$ be any prefix code. Then
\begin{equation*}
\sup_{P\in\mathcal{P}(\varepsilon,k)}\dfrac{A_{P}(L)}{\max\{1,H(P)\}}
=\sup_{2\leq k\in\mathbb{N}}\max \left\{A_{P_{\varepsilon_k,k}}(L), \dfrac{A_{U_k}(L)}{H(U_k)} \right\}.
\end{equation*}
\end{corollary}
\begin{proof}
By Lemma~\ref{lemma5}, we obtain
\[
\begin{aligned}
 \sup_{P\in\mathcal{P}(\varepsilon,k)}\dfrac{A_{P}(L)}{\max\{1,H(P)\}} & =
 \sup_{\substack{2\leq k\in\mathbb{N}\\ 0<\varepsilon\leq(k-1)/k} }\frac{A_{P_{\varepsilon,k}}(L)}{\max\{1,H(P_{\varepsilon,k})\}}  \\
 &=\sup_{2\leq k\in\mathbb{N}}\max \left\{(1-\varepsilon_k)L(1)+\varepsilon_k\widetilde{a}_k, \frac{a_k}{\log_2 k} \right\} \\
  &=\sup_{2\leq k\in\mathbb{N}}\max \left\{A_{P_{\varepsilon_k,k}}(L), \dfrac{A_{U_k}(L)}{H(U_k)} \right\}.
 \end{aligned}
\]
\end{proof}
Corollary~\ref{cor2} shows that, for any prefix code $\mathcal{C}$, 
at least one of the distributions in $\{P_{\varepsilon_k,k},U_k \mid 2\leq k\in\mathbb{N}\}$ is least favorable for $\mathcal{C}$ within the family \(\mathcal{P}(\varepsilon,k)\). Let
\begin{equation} \label{eq54}
\begin{aligned}
T(\mathcal{C}) & \triangleq \sup_{P\in\{P_{\varepsilon_k,k},U_k \mid 2\leq k\in\mathbb{N}\}}\dfrac{A_{P}(L_\mathcal{C})}{\max\{1,H(P)\}}  \\
& =\sup_{2\leq k\in\mathbb{N}}\max \left\{A_{P_{\varepsilon_k,k}}(L_\mathcal{C}), \dfrac{A_{U_k}(L_\mathcal{C})}{H(U_k)} \right\}    \\
& = \sup_{2\leq k\in\mathbb{N}}  \max \left\{(1-\varepsilon_k)L_{\mathcal{C}}(1)+\varepsilon_k\widetilde{a}_k, \frac{a_k}{\log_2 k} \right\} \\
\end{aligned}
\end{equation}
Therefore, from Theorem~\ref{thm3} and Corollary~\ref{cor2}, we immediately obtain the following theorem.
\begin{theorem} \label{thm4}
Let $\mathcal{C}$ be any prefix code. Then
\begin{equation*}
 \sup_{P\in\mathcal{P}_\mathbb{N}}\dfrac{A_{P}(L_\mathcal{C})}{\max\{1,H(P)\}} = \sup_{2\leq k\in\mathbb{N}}\max \left\{A_{P_{\varepsilon_k,k}}(L_\mathcal{C}), \dfrac{A_{U_k}(L_\mathcal{C})}{H(U_k)} \right\};
\end{equation*}
that is, $C_{\mathcal{C}}^{*}=T(\mathcal{C})$.
\end{theorem}
\begin{proof}
By Theorem~\ref{thm3} and Corollary~\ref{cor2}, we have
\begin{equation*}
\begin{aligned}
C_{\mathcal{C}}^{*}&=\sup_{P\in\mathcal{P}_\mathbb{N}}\dfrac{A_{P}(L_\mathcal{C})}{\max\{1,H(P)\}} = \sup_{P\in\mathcal{P}(\varepsilon,k)}\dfrac{A_{P}(L_\mathcal{C})}{\max\{1,H(P)\}}   \\
	& =\sup_{2\leq k\in\mathbb{N}}\max \left\{A_{P_{\varepsilon_k,k}}(L_\mathcal{C}), \dfrac{A_{U_k}(L_\mathcal{C})}{H(U_k)} \right\} =T(\mathcal{C}).
\end{aligned}	
\end{equation*}
\end{proof}
Theorem~\ref{thm4} shows that, for any prefix code $\mathcal{C}$, 
at least one of the distributions in $\{P_{\varepsilon_k,k},U_k \mid 2\leq k\in\mathbb{N}\}$ 
is least favorable for $\mathcal{C}$ within the family \(\mathcal{P}_\mathbb{N}\).
Therefore, when considering optimal UCI, it suffices to consider the family of distributions $\{P_{\varepsilon_k,k},U_k \mid 2\leq k\in\mathbb{N}\}$. This is a very interesting result. 

We give two proofs of Theorem~\ref{thm4}: the proof above, via
Theorem~\ref{thm3} and Corollary~\ref{cor2}, and a direct proof in Appendix~\ref{another}. 
The former is natural, as it reflects the complete
process of studying the problem and discovering the conclusion of Theorem~\ref{thm4}; the latter is shorter, at the price that the statement of the theorem has to be guessed in advance. Lemma~\ref{lemma2}, which arose in the
course of solving the problem, is stated and proved in this section and is used in both proofs.

\section{The Optimal UCI and the UCI Inequality}\label{Inequality}
In this section, we construct an optimal UCI and show that its minimum expansion factor satisfies 
\[
2.0001247570361019786170117629757988625444 \le C^* \le 2.0001247570361019786170117629757988625463.
\] 
To this end, we establish an important inequality in UCI, which plays a role analogous to that of the Kraft inequality~\cite{kraft} for prefix codes.

Yan and Lin~\cite{YL22} proved that the length of the first codeword of optimal UCI is $1$.
Therefore, when considering the optimal UCI, we assume, by default, that the prefix code satisfies $L(1)=1$. This section proves that, for any real number $B\ge1$,
\[
     \mathcal{C} \text{ is a prefix code satisfying } L_{\mathcal{C}}(1)=1 \text{ and }  C_{\mathcal{C}}^*\le B \iff \Sigma(B)\le \frac{7}{4},
\]
where $\Sigma(B)$ is defined in Equation~\eqref{Sigma} below.
The inequality $\Sigma(B)\le \frac{7}{4}$ is a necessary and sufficient condition
for the existence of a UCI code $\mathcal{C}$ with \(L_{\mathcal{C}}(1)=1\) and with minimum expansion factor $C_{\mathcal{C}}^*$ bounded above by \(B\).
This is a strong result, and hence we refer to $\Sigma(B)\le \frac{7}{4}$ as the UCI inequality.
Note that, since \(C_{\mathcal{C}}^*\le B\) implies that \(\mathcal{C}\) is a UCI, we omit the condition that \(\mathcal{C}\) is a UCI in the subsequent theorem statements.

\subsection{Sufficient condition for the UCI inequality}
In this subsection, we first present three auxiliary lemmas, and then prove a sufficient condition for the UCI inequality.
Before stating the lemmas, we introduce some notation.

Let $S(k)\triangleq \sum_{n=1}^{k} L(n)$ denote the sum of the lengths of the first
$k$ codewords. Let $N_n\triangleq |\{m\in\mathbb{N}\mid L(m)=n\}$ denote the number of codewords of length $n$, and let $\mathcal{A}_n\triangleq \sum_{j=1}^{n} N_j$ denote the number of codewords of length at most \(n\).
Define \(\mathcal{B}_n\triangleq \sum_{j=1}^{n} \mathcal{A}_j\). In particular, define \(\mathcal{A}_0=\mathcal{B}_0=0\).
For any prefix code \(\mathcal{C}\) and \(B \in \mathbb{R}\), let
\begin{equation}\label{eq55}
\begin{aligned}
 c_{B,\mathcal{C}}(k) & \triangleq \min \Big\{L_{\mathcal{C}}(1)+(k-1)\big[L_{\mathcal{C}}(1)+\big(B-L_{\mathcal{C}}(1)\big)\sigma_k\big],Bk\log_2 k\Big\},  \\
   d_{n,\mathcal{C}}(B)  & \triangleq \sup_{2\le k \in \mathbb{N}} \{nk- c_{B,\mathcal{C}}(k)\}   \\
\end{aligned}
\end{equation}
for all $2\le k \in \mathbb{N}$ and $n\in \mathbb{N}$.
When the prefix code \(C\) satisfies \(L_{\mathcal{C}}(1)=1\),  the above notation is simplified to
\begin{equation}\label{eq56}
\begin{aligned}
	c_{B}(k) & \triangleq \min \Big\{1+(k-1)\big[1+(B-1)\sigma_k\big],Bk\log_2 k\Big\},  \\
	d_{n}(B)  & \triangleq \sup_{2\le k \in \mathbb{N}} \{nk- c_{B}(k)\}.   \\
\end{aligned}
\end{equation}
For any \(B \in \mathbb{R}\), define
\begin{equation} \label{Sigma}
  \Sigma(B) \triangleq \sum_{n=3}^\infty \lceil d_{n}(B) \rceil 2^{-n}, \qquad
  \Sigma_N(B) \triangleq \sum_{n=3}^N \lceil d_{n}(B) \rceil 2^{-n}
\end{equation}
for all $3\le N\in  \mathbb{N}$. We now prove three auxiliary lemmas.
\begin{lemma} \label{lemma6}
For any prefix code \(\mathcal{C}\) and \(B \in \mathbb{R}\), then $C_{\mathcal{C}}^{*}\le B$ if and only if \(S(k)\le c_{B,\mathcal{C}}(k)\) for all \(2\le k \in \mathbb{N}\).\footnote{Given that the prefix code \(\mathcal{C}\) satisfies \(L_{\mathcal{C}}(1)=1\), the conclusion of Lemma~\ref{lemma6} is revised to $C_{\mathcal{C}}^{*}\le B$ if and only if \(S(k)\le c_{B}(k)\) for all \(2\le k \in \mathbb{N}\). A similar situation holds for Lemma~\ref{lemma7}.}
\end{lemma}
\begin{proof}
By Theorem~\ref{thm4} and the definition of \(T(\mathcal{C})\), it follows that
\[
\begin{aligned}
C_{\mathcal{C}}^{*}\le B &  \iff T(\mathcal{C})\le B    \\
                         &  \iff \sup_{2\leq k\in\mathbb{N}}  \max \left\{(1-\varepsilon_k)L_{\mathcal{C}}(1)+\varepsilon_k\widetilde{a}_k, \frac{a_k}{\log_2 k} \right\} \le B   \\
                &  \iff \begin{cases}
                         	(1-\varepsilon_k)L_{\mathcal{C}}(1)+\varepsilon_k\widetilde{a}_k \le B,\quad \forall \ 2\le k\in\mathbb{N} \\
                         	\dfrac{a_k}{\log_2 k} \le B,\quad \forall \ 2\le k\in\mathbb{N}
                    \end{cases}   \\
                &  \iff \begin{cases}
                	(1-\varepsilon_k)L_{\mathcal{C}}(1)+\varepsilon_k \dfrac{S(k)-L_{\mathcal{C}}(1)}{k-1} \le B,\quad \forall \ 2\le k\in\mathbb{N} \\
                	\dfrac{S(k)}{k\log_2 k} \le B,\quad \forall \ 2\le k\in\mathbb{N}
                \end{cases}   \\
                & \iff   \begin{cases}
                S(k)\le L_{\mathcal{C}}(1)+(k-1)\big[L_{\mathcal{C}}(1)+\big(B-L_{\mathcal{C}}(1)\big)\sigma_k\big],\quad \forall \ 2\le k\in\mathbb{N} \\
                	S(k) \le Bk\log_2 k,\quad \forall \ 2\le k\in\mathbb{N}
                \end{cases}   \\
                & \iff S(k)\le c_{B,\mathcal{C}}(k) ,\quad \forall \ 2\le k\in\mathbb{N}.
\end{aligned}
\]
\end{proof}
\begin{lemma} \label{lemma7}
\begin{enumerate}
	\item[$(1)$] For any positive integer \(k\),
\[
   S(k)= \max_{n\in\mathbb{N}} \big\{nk -\mathcal{B}_{n-1} \big\}.
\]
	\item[$(2)$] If \(S(k)\le c_{B,\mathcal{C}}(k)\) for all \(2\le k \in \mathbb{N}\), then $\mathcal{B}_n \ge \lceil d_{n+1,\mathcal{C}}(B)\rceil$ for all integer $n\ge0$.
\end{enumerate}
\end{lemma}
\begin{proof}
\begin{enumerate}
\item[$(1)$] By interchanging the order of summation, we compute \(S(k)\) as follows.
\[
\begin{aligned}
S(k)&=\sum_{m=1}^k L(m)=\sum_{m=1}^k \sum_{n=1}^{L(m)}1    \\
    &=\sum_{m=1}^k \sum_{n=1}^\infty \mathbf{1}_{\{x\mid L(m)\ge x\}}(n)                      \\
    &=\sum_{n=1}^\infty \sum_{m=1}^k \mathbf{1}_{\{x\mid L(m)\ge x\}}(n)  \\
    &=\sum_{n=1}^\infty \big|\{m\le k\mid L(m) \ge n \} \big|,    \\
\end{aligned}
\]
where $\mathbf{1}_A(x)$ denotes the indicator function.
Since $\big|\{m\le k\mid L(m) \ge n \} \big|$ denotes the number of codewords of length at least \(n\) among the first \(k\) codewords, we obtain
\[
\begin{aligned}
	S(k)&=\sum_{n=1}^\infty \big|\{m\le k\mid L(m) \ge n \} \big|   \\
	    &=\sum_{n=1}^\infty \max\big\{k-\mathcal{A}_{n-1},0\big\}   \\
	    &\overset{(a)}{=}\max_{n\in \mathbb{N}} \sum_{j=1}^n (k-\mathcal{A}_{j-1})     \\
	    &=\max_{n\in \mathbb{N}}\big\{nk -\mathcal{B}_{n-1} \big\},
\end{aligned}
\]
where $(a)$ holds because in the sequence \(\left\{a_n=\max\big\{k-\mathcal{A}_{n-1},0\big\} \right\}_{n=1}^\infty\), only finitely many initial terms are greater than \(0\), and all subsequent terms are identically zero.
\item[$(2)$] For any $n\in\mathbb{N}$ and any \(2\le k\in\mathbb{N}\), it follows that
\[
 nk -\mathcal{B}_{n-1} \le S(k)  \le c_{B,\mathcal{C}}(k),
\]
and hence we obtain $\mathcal{B}_{n-1} \ge nk-c_{B,\mathcal{C}}(k)$ for all $n\in\mathbb{N}$ and \(2\le k\in\mathbb{N}\).
Therefore, we have
$$\mathcal{B}_{n-1} \ge   d_{n,\mathcal{C}}(B) =\sup_{2\le k \in \mathbb{N}} \{nk- c_{B,\mathcal{C}}(k)\} $$
for all $n\in\mathbb{N}$.
Also, because \(\mathcal{B}_n\) is an integer, it follows that $\mathcal{B}_n \ge \lceil d_{n+1,\mathcal{C}}(B)\rceil$ for all integer $n\ge0$.
\end{enumerate}
\end{proof}
The final auxiliary lemma uses the previously defined notation to express the Kraft sum, i.e., \(\sum_{n=1}^\infty 2^{-L(n)}\).
\begin{lemma} \label{lemma8}
The following identity holds:
\begin{equation*}
\sum_{n=1}^\infty 2^{-L(n)} = \frac{1}{2} \sum_{n=1}^\infty \mathcal{A}_n 2^{-n}= \frac{1}{4} \sum_{n=1}^\infty \mathcal{B}_n 2^{-n}.
\end{equation*}	
\end{lemma}
\begin{proof}
Since \(N_n =\mathcal{A}_n-\mathcal{A}_{n-1}\), it follows that
\[
 \begin{aligned}
 \sum_{n=1}^\infty 2^{-L(n)}&= \sum_{n=1}^\infty N_n 2^{-n}  \\
                            &= \sum_{n=1}^\infty \big(\mathcal{A}_n-\mathcal{A}_{n-1}\big)\ 2^{-n}  \\
                            &=\sum_{n=1}^\infty\mathcal{A}_n2^{-n}-\sum_{n=1}^\infty\mathcal{A}_n2^{-(n+1)}  \\
                            &=\frac{1}{2} \sum_{n=1}^\infty \mathcal{A}_n 2^{-n}.
 \end{aligned}
\]
Because $\mathcal{A}_n=\mathcal{B}_n-\mathcal{B}_{n-1}$, we obtain
\[
\begin{aligned}
	\sum_{n=1}^\infty 2^{-L(n)}&= \frac{1}{2} \sum_{n=1}^\infty \mathcal{A}_n 2^{-n}  \\
	&=\frac{1}{2} \sum_{n=1}^\infty \big(\mathcal{B}_n-\mathcal{B}_{n-1}\big)\ 2^{-n}  \\
	&=\frac{1}{4} \sum_{n=1}^\infty \mathcal{B}_n 2^{-n}.
\end{aligned}
\]
\end{proof}
We now propose and prove a sufficient condition for the UCI inequality.
\begin{theorem} \label{thm5}
For any real number \(B\),\footnote{From Corollary~\ref{cor1}, we obtain $B\ge C_{\mathcal{C}}^{*}\ge 2$. Therefore, we omit the condition that $B\ge1$.} if there exists a prefix code \(\mathcal{C}\) such that \(L_{\mathcal{C}}(1) = 1\) and $C_{\mathcal{C}}^{*}\le B$, then \(\Sigma(B)\le \frac{7}{4}\).
\end{theorem}
\begin{proof}
Since \(L_{\mathcal{C}}(1) = 1\) and $C_{\mathcal{C}}^{*}\le B$, it follows from Lemma~\ref{lemma6} that $S(k)\le c_{B}(k)$ for all $2\le k\in \mathbb{N}$.
From Lemma~\ref{lemma7}, we have $\mathcal{B}_n \ge \lceil d_{n+1}(B)\rceil$ for all integer $n\ge0$.
Because $\mathcal{C}$ is a prefix code, we obtain \(N_1 = 0\) or \(N_1 = 1\).
Moreover, since \(L_{\mathcal{C}}(1) = 1\), it follows that \(N_1 = 1\).
From the definitions of \(\mathcal{A}_n\) and \(\mathcal{B}_n\),
we have \(\mathcal{B}_1=\mathcal{A}_1=N_1=1\).
By Lemma~\ref{lemma8}, it follows that
\[
\begin{aligned}
 \sum_{n=1}^\infty 2^{-L(n)} &= \frac{1}{4} \sum_{n=1}^\infty \mathcal{B}_n 2^{-n}   \\
                             &= \frac{1}{8}+\frac{1}{4}\sum_{n=2}^\infty \mathcal{B}_n 2^{-n}  \\
                             &\ge \frac{1}{8}+\frac{1}{4}\sum_{n=2}^\infty \lceil d_{n+1}(B)\rceil 2^{-n}  \\
                             &= \frac{1}{8}+\frac{1}{2}\sum_{n=3}^\infty \lceil d_{n}(B)\rceil 2^{-n}  \\
                              &= \frac{1}{8}+\frac{1}{2}\Sigma(B). \\
\end{aligned}
\]
From the Kraft inequality~\cite{kraft}, we obtain
\[
 \frac{1}{8}+\frac{1}{2}\Sigma(B) \le \sum_{n=1}^\infty 2^{-L(n)} \le 1,
\]
and hence \(\Sigma(B)\le \frac{7}{4} \).
\end{proof}
\subsection{Necessary condition for the UCI inequality}
In this subsection, we prove a necessary condition for the UCI inequality.
That is, assuming inequality \(\Sigma(B)\le \frac{7}{4} \) holds, it suffices to construct a prefix code \(\mathcal{C}\) such that \(L_{\mathcal{C}}(1)=1\) and $C_{\mathcal{C}}^{*}\le B$.
Therefore, we first construct a code \(\mathcal{C}(B)\) for real number \(B\).
Next, we explain the rationale for constructing \(\mathcal{C}(B)\). Finally, we prove that the constructed \(\mathcal{C}(B)\) satisfies the required conditions.
\begin{construction} \label{con1}
For each \(B\in\mathbb{R}\), construct a code \(\mathcal{C}(B)\) as follows.
Let the code \(\mathcal{C}(B)\) satisfy $\mathcal{B}_1=1$ and $\mathcal{B}_n=\lceil d_{n+1}(B)\rceil$ for all $2\le n\in\mathbb{N}$.
\end{construction}
Here, we provide three explanations regarding the validity of Construction~\ref{con1}.

\textbf{First explanation: the range of $B$ for which $\mathcal{C}(B)$ is a valid code.}
We adopt the convention that the codewords are arranged in nondecreasing order of length; that is, \(L(n)\le L(n+1)\) for every positive integer \(n\). Therefore, once the sequence \(\{N_n\}_{n=1}^{\infty}\) is determined, the sequence \(\{L(n)\}_{n=1}^{\infty}\) is uniquely determined as well.

Construction~\ref{con1} fixes the sequence \(\{\mathcal{B}_n\}_{n=1}^{\infty}\). Since \(\mathcal{A}_n=\mathcal{B}_n-\mathcal{B}_{n-1}\) and \(N_n=\mathcal{A}_n-\mathcal{A}_{n-1}\), with the convention \(\mathcal{B}_0=\mathcal{A}_0=0\), the sequence \(\{N_n\}_{n=1}^{\infty}\) is determined as well, and in fact
\[
N_1=1,\qquad N_n=\mathcal{B}_n-2\mathcal{B}_{n-1}+\mathcal{B}_{n-2}\quad(n\ge2).
\]
The construction therefore defines a code precisely when all \(N_n\) are nonnegative. This motivates the following definition.

\begin{definition}\label{def:valid}
A code \(\mathcal{C}\) is said to be \emph{valid} if \(N_n\ge0\) for every positive integer \(n\), where $N_n$ denotes the prescribed number of codewords of
$\mathcal{C}$ having length $n$; equivalently, if there exists a nondecreasing sequence \(\{L(n)\}_{n=1}^{\infty}\) of positive integers containing exactly \(N_n\) terms equal to \(n\) for every \(n\ge1\), in which case
$L_{\mathcal{C}}(m)=L(m)$ for all $m\ge1$. 
A valid code \(\mathcal{C}\) is called a \emph{prefix code} if it additionally satisfies the Kraft inequality
	\begin{equation}\label{eq:kraft}
		\sum_{n\ge1}N_n2^{-n}\ \le\ 1.
	\end{equation}
\end{definition}

By Definition~\ref{def:valid}, \(\mathcal{C}(B)\) is a prefix code if and only if it is valid and Equation~\eqref{eq:kraft} holds. The remainder of this first explanation is devoted to the range of \(B\) for which \(\mathcal{C}(B)\) is a valid code; the Kraft inequality will be verified in the proof of Theorem~\ref{thm6}.
Next, we prove several lemmas.

\begin{lemma}\label{lem:log}
	For every real number \(x\ge2\),
	\[
	x\log_2x\ \ge\ 2(x-1).
	\]
\end{lemma}
\begin{proof}
	Let \(g_1(x)\triangleq\log_2x-2+\frac{2}{x}\); the asserted inequality is equivalent to \(g_1(x)\ge0\). Since \(g_1(2)=1-2+1=0\) and
	\[
	g_1'(x)=\frac{1}{x\ln2}-\frac{2}{x^2}=\frac{x-2\ln2}{x^2\ln2}>0\qquad(x\ge2>2\ln2),
	\]
	the function \(g_1(x)\) is strictly increasing on \([2,\infty)\), and hence \(g_1(x)\ge g_1(2)=0\).
\end{proof}

\begin{lemma}\label{lem:max}
	Let \(B>1\) and \(n\in\mathbb{N}\). Then \(d_n(B)<\infty\), and there exists an integer
	\(k_n^*\ge2\) such that
	\[
	d_n(B)=nk_n^*-c_B(k_n^*).
	\]
\end{lemma}
\begin{proof}
We show that, for every \(k\ge2\), both numbers in the minimum
\[
c_B(k)=\min\Big\{1+(k-1)\big[1+(B-1)\sigma_k\big],\ Bk\log_2k\Big\}
\]
are strictly larger than \((B-1)(k-1)\log_2(k-1)\).

For the first number, since \(B-1>0\), \(k-1\ge1\), and
\(\sigma_k\ge2+\log_2(k-1)>\log_2(k-1)\) (see Lemma~\ref{lemma:sigma}(2)),
\[
1+(k-1)\big[1+(B-1)\sigma_k\big]
=k+(k-1)(B-1)\sigma_k
>(k-1)(B-1)\log_2(k-1).
\]
For the second number, since \(B>B-1>0\), \(k\log_2k>0\), and \(x\log_2x\) is strictly
increasing on \([1,\infty)\) with \(k>k-1\ge1\), we obtain
\[
Bk\log_2k>(B-1)k\log_2k>(B-1)(k-1)\log_2(k-1)
\]
for all \(k\ge2\). Hence \(c_B(k)>(B-1)(k-1)\log_2(k-1)\) for all \(k\ge2\).

Consequently, for every \(k\ge2\),
\[
nk-c_B(k)
<nk-(B-1)(k-1)\log_2(k-1)
=k\Big[n-(B-1)\frac{k-1}{k}\log_2(k-1)\Big].
\]
Since \(\frac{k-1}{k}\to1\) and \(\log_2(k-1)\to\infty\), the bracket tends to \(-\infty\);
as \(k\ge2\), the right-hand side tends to \(-\infty\).

Fix an integer \(k_0\ge2\). Since \(\lim_{k\to\infty}\big(nk-c_B(k)\big)=-\infty\), there exists
\(k_1>k_0\) such that \(nk-c_B(k)<nk_0-c_B(k_0)\) for every \(k\ge k_1\). Hence every value
with \(k\ge k_1\) is smaller than \(nk_0-c_B(k_0)\), and, since \(k_0<k_1\), the latter is itself
one of the values in the maximum below. Therefore
\[
d_n(B)=\sup_{k\ge2}\big\{nk-c_B(k)\big\}=\max_{2\le k<k_1}\big\{nk-c_B(k)\big\}<\infty .
\]
The maximum is over the nonempty finite set \(\{2,3,\ldots,k_1-1\}\), so it is attained
at some integer \(k_n^*\ge2\).
\end{proof}
\begin{lemma}\label{lem:piece}
	Let \(\{N_n\}\) be defined by Construction~\ref{con1}. Then, for every positive integer \(n\),
	\begin{equation}\label{eq:piecewise}
		N_n=
		\begin{cases}
			1, & n=1,\\[2pt]
			\big\lceil d_3(B)\big\rceil-2, & n=2,\\[2pt]
			\big\lceil d_4(B)\big\rceil-2\big\lceil d_3(B)\big\rceil+1, & n=3,\\[2pt]
			\big\lceil d_{n+1}(B)\big\rceil+\big\lceil d_{n-1}(B)\big\rceil-2\big\lceil d_n(B)\big\rceil, & n\ge4.
		\end{cases}
	\end{equation}
\end{lemma}
\begin{proof}
	By Construction~\ref{con1}, \(\mathcal{B}_1=1\) and \(\mathcal{B}_j=\lceil d_{j+1}(B)\rceil\) for every
	\(j\ge2\); moreover \(\mathcal{B}_0=\mathcal{A}_0=0\) by convention. Since
	\[
	N_1=\mathcal{A}_1-\mathcal{A}_0=\mathcal{B}_1-\mathcal{B}_0=1,
	\]
	and since \(N_n=\mathcal{B}_n-2\mathcal{B}_{n-1}+\mathcal{B}_{n-2}\) for every \(n\ge2\),
	substitution gives
	\[
	N_2=\mathcal{B}_2-2\mathcal{B}_1+\mathcal{B}_0=\lceil d_3(B)\rceil-2,
	\qquad
	N_3=\mathcal{B}_3-2\mathcal{B}_2+\mathcal{B}_1=\lceil d_4(B)\rceil-2\lceil d_3(B)\rceil+1 .
	\]
	For \(n\ge4\), the three indices \(n\), \(n-1\), and \(n-2\) are all at least \(2\), so that
	\[
	N_n=\big\lceil d_{n+1}(B)\big\rceil+\big\lceil d_{n-1}(B)\big\rceil-2\big\lceil d_n(B)\big\rceil .
	\]
\end{proof}
\begin{lemma}\label{lem:aux}
\begin{enumerate}
		\item[$(1)$] Let \(B>1\). Then \(d_4(B)\ge d_3(B)+2\).
		\item[$(2)$] Let \(B>1\) and let \(n\ge4\). Then \(N_n\ge0\).
		\item[$(3)$] Let \(B\in[1.5,\,2.5)\). Then $1<d_3(B)\le3 $.
\end{enumerate}
\end{lemma}
\begin{proof}
	\begin{enumerate}
		\item[$(1)$]
		Let \(k=k_3^*\) be the integer given by Lemma~\ref{lem:max} for \(n=3\), so that
		\(d_3(B)=3k-c_B(k)\). Since \(k\) is admissible in the supremum defining \(d_4(B)\),
		\[
		d_4(B)\ \ge\ 4k-c_B(k)=\big(3k-c_B(k)\big)+k=d_3(B)+k\ \ge\ d_3(B)+2,
		\]
	where the last inequality uses \(k\ge2\).
		\item[$(2)$]
	Let \(k_n^*\) be the integer given by Lemma~\ref{lem:max}.
	From Lemmas~\ref{lem:max} and~\ref{lem:piece}, we obtain
	\[
	\begin{aligned}
		N_n  & =\big\lceil d_{n+1}(B)\big\rceil+\big\lceil d_{n-1}(B)\big\rceil-2\big\lceil d_n(B)\big\rceil  \\
		& \ge \big\lceil (n+1)k_n^*-c_B(k_n^*)\big\rceil+\big\lceil (n-1)k_n^*-c_B(k_n^*)\big\rceil-2\big\lceil nk_n^*-c_B(k_n^*)\big\rceil  \\
		&=(n+1)k_n^*+(n-1)k_n^*-2nk_n^*+\big\lceil -c_B(k_n^*)\big\rceil+\big\lceil -c_B(k_n^*)\big\rceil-2\big\lceil -c_B(k_n^*)\big\rceil  \\
		&=0.
	\end{aligned}
	\]
		\item[$(3)$]
		\emph{Lower bound.} Since \(\sigma_2=2\), the two numbers in the minimum defining \(c_B\)
		both equal \(2B\) at \(k=2\), so \(c_B(2)=2B\) and
		\[
		d_3(B)\ \ge\ 3\cdot2-c_B(2)=6-2B>1
		\]
		for all \(B\in[1.5,\,2.5)\).
		
\emph{Upper bound.} Since both numbers in the minimum defining \(c_B\) are increasing in \(B\), 
it suffices to prove \(c_B(k)\ge3k-3\) for every \(k\ge2\) at \(B=1.5\); then \(3k-c_B(k)\le3\) for every \(k\ge2\), whence \(d_3(B)\le3\). 
We now consider \(B=1.5\).
When \(k=2\), we have
\[
 c_{1.5}(2)=\min\Big\{1+\big(1+0.5\sigma_2\big),\ 1.5\times 2\log_2 2\Big\}=3,
\]
so \(c_{1.5}(k)\ge3k-3\) holds for \(k=2\).
When \(k\ge3\), Lemma~\ref{lem:log} gives
		\[
		1.5\times k\log_2k\ \ge\ 3(k-1)=3k-3 ,
		\]
		and
		\[
		1+(k-1)\Big(1+0.5\sigma_k\Big)\ \ge\ 3k-3
		\iff
		\sigma_k\ \ge\ \frac{4k-6}{k-1}=4-\frac{2}{k-1},
		\]
		which follows from \(\sigma_k\ge2+\log_2(k-1)\)  (see Lemma~\ref{lemma:sigma}(2)) together with
		\[
		\log_2(k-1)\ \ge\ 2-\frac{2}{k-1} ;
		\]
		the latter holds with equality for \(k=3\), by direct verification for \(k=4\), and for
		\(k\ge5\) because \(\log_2(k-1)\ge2\ge2-\frac{2}{k-1}\).
	\end{enumerate}
\end{proof}
With the explicit form of \(N_n\) in Equation~\eqref{eq:piecewise} and the estimates collected in Lemma~\ref{lem:aux}, we are now in a position to prove that Construction~\ref{con1} is valid on \([1.5,2.5)\).
\begin{theorem}\label{thm:valid}
	Let \(B\in[1.5,\,2.5)\). Then the code \(\mathcal{C}(B)\) of Construction~\ref{con1} is valid;
	that is, \(N_n\ge0\) for every positive integer \(n\).
\end{theorem}
\begin{proof}
By Equation~\eqref{eq:piecewise}, \(N_1=1\).
By Lemma~\ref{lem:aux}$(3)$, \(1<d_3(B)\le3\), so that \(\lceil d_3(B)\rceil\in\{2,3\}\); hence, by Equation~\eqref{eq:piecewise},
\[
N_2=\lceil d_3(B)\rceil-2\ \ge\ 0 .
\]
By Lemma~\ref{lem:aux}$(1)$, \(d_4(B)\ge d_3(B)+2\), hence \(\lceil d_4(B)\rceil\ge\lceil d_3(B)+2\rceil=\lceil d_3(B)\rceil+2\).
Again by Equation~\eqref{eq:piecewise},
\[
\begin{aligned}
N_3 & =\lceil d_4(B)\rceil-2\lceil d_3(B)\rceil+1 \\
      &  \ge \lceil d_3(B)\rceil+2 -2\lceil d_3(B)\rceil+1 \\
     & =3-\lceil d_3(B)\rceil\\
     & \ge 0 ,
\end{aligned}
\]
where the last inequality uses \(\lceil d_3(B)\rceil\le3\).
Finally, since \(B>1\), Lemma~\ref{lem:aux}$(2)$ gives \(N_n\ge0\) for every \(n\ge4\). 
\end{proof}
\begin{remark}\label{rem:range}
For completeness, we record that Construction~\ref{con1} produces a valid code if and only if
\[
B\in(1,\,B_0)\cup[1.5,\,2.5),
\qquad
B_0=1+\frac{9}{4\sigma_5}=1.3837052\cdots .
\]
Moreover, since \(\Sigma(B)>\frac74\) for every \(B<1.5\) (see the proof of
Theorem~\ref{thm6}), the code \(\mathcal{C}(B)\) is a prefix code if and only if
\(B\in[1.5,2.5)\) and \(\Sigma(B)\le\frac74\). 
The proof of the above sufficient and necessary condition is lengthy, and we omit it here: for \(B<1.5\) the code \(\mathcal{C}(B)\) is never a prefix code, so this
part of the range is irrelevant to the UCI inequality, and the proof of Theorem~\ref{thm6}
below uses only \(B\in[1.5,2.5)\). 
We also note that for \(B\le1\) Construction~\ref{con1} is not defined at all, since
\(d_n(B)=+\infty\) for every \(n\ge2\).
\end{remark}
\textbf{Second explanation: from the length sequence to an actual prefix code.}
In principle, Construction~\ref{con1} always specifies a code: once the sequence \(\{L(n)\}_{n=1}^\infty\) is determined, the explicit codewords can be derived from alphabetic codes~\cite{91NN,91Yeung,2024code}; for an explicit construction we also refer to~\cite{Yan26}. What has to be verified is that this code is a prefix code, which by Definition~\ref{def:valid} amounts to two properties, namely validity and the Kraft inequality. Theorem~\ref{thm:valid} supplies the first property whenever \(B\in[1.5,2.5)\). As for the second, it follows from Lemma~\ref{lemma8} that the Kraft sum of the length sequence of Construction~\ref{con1} equals
\begin{equation*}
	\sum_{n=1}^\infty N_n2^{-n}=\frac{1}{4}\sum_{n=1}^\infty\mathcal{B}_n2^{-n}
	=\frac{1}{8}+\frac{1}{2}\Sigma(B),
\end{equation*}
so that Equation~\eqref{eq:kraft} is equivalent to \(\Sigma(B)\le\frac74\). Consequently, for every \(B\in[1.5,2.5)\) satisfying \(\Sigma(B)\le\frac74\), the code \(\mathcal{C}(B)\) is a prefix code with \(L_{\mathcal{C}(B)}(1)=1\).

\textbf{Finally explanation: the rationale for this construction.} 
Since an optimal UCI must have $L(1)=1$~\cite{YL22}, setting
$\mathcal{B}_1=1$ gives $N_1=\mathcal{A}_1=\mathcal{B}_1-\mathcal{B}_0=1$, and hence \(L_{\mathcal{C}(B)}(1) = 1\). 
This is the reason for the exceptional value \(\mathcal{B}_1=1\). 
Were we to set \(\mathcal{B}_1=\lceil d_2(B)\rceil\) instead, 
then for \(B=2.1\) we would obtain
\[
\mathcal{B}_1=\big\lceil d_2(2.1)\big\rceil=\lceil-0.2\rceil=0,
\]
whence \(N_1=\mathcal{A}_1=\mathcal{B}_1=0\) and \(L_{\mathcal{C}(B)}(1)\ge2\),
so that the resulting code would not be an optimal UCI.

For \(n\ge2\), the value of \(\mathcal{B}_n\) is forced. 
On the one hand, for the Kraft inequality in Equation~\eqref{eq12} below to hold, 
one needs \(\mathcal{B}_n\le\lceil d_{n+1}(B)\rceil\) for every \(n\ge2\); 
on the other hand, for the estimate in Equation~\eqref{eq13} below to hold, 
one needs \(\mathcal{B}_n\ge\lceil d_{n+1}(B)\rceil\) for every \(n\ge2\).  
Therefore, one must set
\[
\mathcal{B}_n=\big\lceil d_{n+1}(B)\big\rceil,
\]
for $2\le n\in\mathbb{N}$, which is exactly the choice made in Construction~\ref{con1}.

Before proving the necessary condition, we need to prove a related lemma.
\begin{lemma} \label{lemma9}
The function
\[
\Sigma : B\longmapsto\Sigma(B)
\]
is nonincreasing on \([1,\infty)\). That is, if $1\le B_1\le B_2$, then $\Sigma(B_1)\ge\Sigma(B_2)$.
\end{lemma}
\begin{proof}
We prove the monotonicity of \(\Sigma\) in two steps.
\begin{enumerate}
\item First, we prove that if $1\le B_1\le B_2$, then $c_{B_1}(k)\le c_{B_2}(k)$ for all
$2\le k\in\mathbb{N}$.

Suppose $1\le B_1\le B_2$.
Since \(k\ge2\) and $\sigma_k\ge2$, we have
\[
\begin{aligned}
1+(k-1)\big[1+(B_1-1)\sigma_k\big] &\le  1+(k-1)\big[1+(B_2-1)\sigma_k\big],  \\
B_1 k\log_2 k & \le B_2 k\log_2 k.  \\
\end{aligned}
\]
Moreover, since \(c_{B}(k) = \min \Big\{1+(k-1)\big[1+(B-1)\sigma_k\big],Bk\log_2 k\Big\}\), we obtain
$c_{B_1}(k)\le c_{B_2}(k)$.
\item Second, we prove that if $1\le B_1\le B_2$, then $\Sigma(B_1)\ge\Sigma(B_2)$.

Suppose $1\le B_1\le B_2$, then $c_{B_1}(k)\le c_{B_2}(k)$ for all
$2\le k\in\mathbb{N}$, and hence $nk-c_{B_1}(k)\ge nk-c_{B_2}(k)$ for all $2\le k\in\mathbb{N}$ and $n\in\mathbb{N}$. Furthermore, by $d_{n}(B)=\sup_{2\le k \in \mathbb{N}} \{nk- c_{B}(k)\}$, we obtain $d_n(B_1)\ge d_n(B_2)$ for all $n\in\mathbb{N}$.
Since $\Sigma(B) \triangleq \sum_{n=3}^\infty \lceil d_{n}(B) \rceil 2^{-n}$, it follows that
$\Sigma(B_1)\ge\Sigma(B_2)$.
\end{enumerate}
\end{proof}

We now prove a necessary condition for the UCI inequality.
\begin{theorem} \label{thm6}
For any real number \(B\ge1\), if \(\Sigma(B)\le \frac{7}{4}\), then there exists a prefix code \(\mathcal{C}\) such that \(L_{\mathcal{C}}(1) = 1\) and $C_{\mathcal{C}}^{*}\le B$.
\end{theorem}
\begin{proof}
We distinguish three cases according to the value of \(B\).	
\begin{enumerate}
\item \emph{Case 1: \(B\ge2.5\).} By Theorem~\ref{thm9}, the \(\nu\) code~\cite{Yan26}
satisfies \(L_{\nu}(1)=1\) and \(C_{\nu}^{*}\le2.0386<B\), so the required
prefix code exists. Moreover, applying Theorem~\ref{thm5} to this code gives
$\Sigma(B)\le\frac74$; the hypothesis is therefore satisfied automatically in
this range, and no separate verification is required.
\item \emph{Case 2: \(1.5\le B<2.5\).} Let \(\mathcal{C}(B)\) be the code of Construction~\ref{con1}.
By Theorem~\ref{thm:valid}, \(\mathcal{C}(B)\) is valid.
From Lemma~\ref{lemma8}, it follows that
\begin{equation} \label{eq12}
	\begin{aligned}
		\sum_{n=1}^\infty 2^{-L(n)} & = \frac{1}{4} \sum_{n=1}^\infty \mathcal{B}_n 2^{-n} \\
		& = \frac{1}{4} \times2^{-1} + \frac{1}{4}\sum_{n=2}^\infty\lceil d_{n+1}(B)\rceil 2^{-n} \\
		&= \frac{1}{8}+\frac{1}{2}\sum_{n=3}^\infty \lceil d_{n}(B)\rceil 2^{-n}  \\
		&= \frac{1}{8}+\frac{1}{2}\Sigma(B) \\
		& \le 1.
	\end{aligned}
\end{equation}
Then \(\mathcal{C}(B)\) is a prefix code.
Since \(\mathcal{B}_1=1\), we have \(N_1 = \mathcal{A}_1= \mathcal{B}_1 = 1\). Hence, we obtain \(L_{\mathcal{C}(B)}(1)=1\).

It remains to show that \(C_{\mathcal{C}(B)}^{*}\le B\).
We first claim that
\begin{equation} \label{eq98}
c_{B}(k)\ge 2k-1
\end{equation}
for all $2\le k\in\mathbb{N}$. Indeed, by Lemma~\ref{lem:log} and \(B\ge1.5\), we have
\[
Bk\log_2k\ \ge\ \frac{3}{2}k\log_2k\ \ge\ 3(k-1)\ \ge\ 2k-1
\]
for all $2\le k\in\mathbb{N}$. 
Moreover, since \(\sigma_k\ge2\) gives
\[
1+(k-1)\big[1+(B-1)\sigma_k\big] \ge 1+(k-1)\big[1+(1.5-1)\times2\big]=2k-1,
\]
and hence $c_{B}(k)\ge 2k-1$ for all $2\le k\in\mathbb{N}$.

Next, we prove that
\begin{equation} \label{eq11}
	nk -\mathcal{B}_{n-1}\le c_{B}(k)  \qquad  \forall \ n\in\mathbb{N}, \ 2\le k\in\mathbb{N}.
\end{equation}
When $n=1$, we have $nk -\mathcal{B}_{n-1}=k<2k-1\le c_{B}(k)$ for all $2\le k\in\mathbb{N}$.
When $n=2$, we obtain $nk -\mathcal{B}_{n-1}=2k-1\le c_{B}(k)$ by Equation~\eqref{eq98}.
When $n\ge3$, Construction~\ref{con1} gives $\mathcal{B}_m=\lceil d_{m+1}(B)\rceil$ for every $m\ge2$,
it follows that
\begin{equation} \label{eq13}
	nk-c_{B}(k)\le d_n(B) \le \lceil d_{n}(B)\rceil = \mathcal{B}_{n-1}
\end{equation}
for all $2\le k\in\mathbb{N}$, and hence
$nk -\mathcal{B}_{n-1}\le c_{B}(k)$ for all $2\le k\in\mathbb{N}$.
Therefore, Equation~\eqref{eq11} holds.
Furthermore, by Lemmas~\ref{lemma6} and~\ref{lemma7}, we obtain
\[
\begin{aligned}
	\text{Equation~}\eqref{eq11} &  \iff S(k)=	\max_{n\in\mathbb{N}} \big\{nk -\mathcal{B}_{n-1} \big\} \le c_{B}(k)  \qquad  \forall \  2\le k\in\mathbb{N}.      \\
	&  \iff C_{\mathcal{C}(B)}^{*}\le B.    \\
\end{aligned}
\]
\item \emph{Case 3: $1\le B< 1.5$.} We show that no $B$ in this range satisfies \(\Sigma(B)\le\frac74\).
Assume, for contradiction, that \(\Sigma(1.5)\le\frac74\). 
Since \(1.5\in[1.5,2.5)\), applying Case~$2$ with $B=1.5$ yields a prefix code \(\mathcal{C}(1.5)\) with
\(C_{\mathcal{C}(1.5)}^{*}\le 1.5<2\), contradicting Corollary~\ref{cor1}.
Therefore, \(\Sigma(1.5)> \frac{7}{4}\). 
By Lemma~\ref{lemma9}, \(\Sigma\) is nonincreasing; it follows that
\[
\Sigma(B)\ \ge\ \Sigma(1.5)\ >\ \frac{7}{4},
\]
for all \(1\le B<1.5\).
Hence, the hypothesis \(\Sigma(B)\le\frac74\) cannot hold.
\end{enumerate}	

This completes the proof.
\end{proof}

\subsection{The optimal UCI with the minimum expansion factor $C^*$}
The following theorem is an immediate consequence of Theorems~\ref{thm5} and~\ref{thm6}.
\begin{theorem}\label{thm7}
For any real number \(B\ge1\), there exists a prefix code \(\mathcal{C}\) such that \(L_{\mathcal{C}}(1) = 1\) and $C_{\mathcal{C}}^{*}\le B$ if and only if \(\Sigma(B)\le \frac{7}{4}\).	
\end{theorem}
We now present a theorem relating the optimal minimum expansion factor $C^*$ to the UCI inequality.
\begin{theorem}\label{thm8}
For any real number \(B\ge1\),
\begin{enumerate}
\item[$(1)$] if \(\Sigma(B)\le \frac{7}{4}\), then $C^*\le B$;
\item[$(2)$] if \(\Sigma(B)> \frac{7}{4}\), then $C^*\ge B$;
\item[$(3)$] The optimal minimum expansion factor $C^*$ satisfies
\begin{equation} \label{eq14}
  C^*=\inf \left\{ B\ge1 \ \Big| \ \Sigma(B)\le \frac{7}{4}  \right\}.
\end{equation}
\end{enumerate}
\end{theorem}
\begin{proof}
\begin{enumerate}
\item[$(1)$] If \(\Sigma(B)\le \frac{7}{4}\), then by Theorem~\ref{thm7} there exists a prefix code \(\mathcal{C}_0\) such that $C_{\mathcal{C}_0}^{*}\le B$, which further yields
\[
C^*=\inf_{\mathcal{C}\in S_{UCI}}C_{\mathcal{C}}^{*} \le C_{\mathcal{C}_0}^{*}\le B.
\]
\item[$(2)$] If \(\Sigma(B)> \frac{7}{4}\), we first prove that \(B < 2.0386\). Since there exists a prefix code \(\nu\) satisfying \(L_{\nu}(1) = 1\) and $C_{\nu}^{*}\le 2.0386$, it follows from Theorem~\ref{thm7} that \(\Sigma(2.0386)\le \frac{7}{4}\).
Therefore, we have \(\Sigma(2.0386)<\Sigma(B)\).
Furthermore, by Lemma~\ref{lemma9}, we obtain \(B < 2.0386\).

From Corollary~\ref{cor1}, there cannot exist a prefix code \(\mathcal{C}\) satisfying \(L_{\mathcal{C}}(1)\ge2\) and $C_{\mathcal{C}}^{*}\le B$; otherwise, we would have
\[
 C_{\mathcal{C}}^{*} \ge 1+L_{\mathcal{C}}(1) \ge 3>2.0386>B,
\]
which contradicts $C_{\mathcal{C}}^{*}\le B$.
Moreover, by Theorem~\ref{thm7}, there cannot exist a prefix code \(\mathcal{C}\) satisfying \(L_{\mathcal{C}}(1)=1\) and $C_{\mathcal{C}}^{*}\le B$; otherwise, we would have  \(\Sigma(B)\le \frac{7}{4}\), which contradicts \(\Sigma(B)> \frac{7}{4}\).
From the above, it follows that when \(\Sigma(B)> \frac{7}{4}\), there is no prefix code \(\mathcal{C}\) satisfying $C_{\mathcal{C}}^{*}\le B$; that is, if \(\Sigma(B)> \frac{7}{4}\), then $C_{\mathcal{C}}^{*}>B$ holds for any prefix code $\mathcal{C}$. Therefore, we obtain
\[
C^*=\inf_{\mathcal{C}\in S_{UCI}}C_{\mathcal{C}}^{*} \ge B.
\]
\item[$(3)$] Since \(\Sigma(2.0386)\le \frac{7}{4}\), it follows that the set $S\triangleq\{ B\ge1 \mid \Sigma(B)\le \frac{7}{4} \}$ is nonempty.
For every \(B \in S\), it follows from $(1)$ that \(C^* \le B\). Hence, \(C^* \le \inf S\).

Next, we prove that \(\inf S \le C^* \).
From $(2)$, it follows that its contrapositive holds, i.e.,
for any real number \(B\ge1\), if $C^*<B$, then \(\Sigma(B)\le\frac{7}{4}\).
By Theorem~\ref{thm9}, \(C^*\ge 2\). It follows that \(B> C^*\ge 2\), and hence \(B \ge 1\). Thus the contrapositive applies. Therefore, if \(C^*<B\), we obtain \(\Sigma(B)\le\frac{7}{4}\), i.e., \(B \in S\).
Since \(B\) is an arbitrary real number greater than \(C^*\), we have shown that the entire interval \((C^*,\infty)\) is contained in \(S\).
Hence, \(\inf S \le \inf (C^*,\infty) = C^*\). Therefore, Equation~\eqref{eq14} holds.
\end{enumerate}
\end{proof}

Theorem~\ref{thm8}$(1)$ and $(2)$ tell us how to further tighten the bound on the optimal minimum expansion factor \(C^*\), and even how to obtain the exact value of \(C^*\).
On the one hand, taking \(B_1 > 2\) with \(\Sigma(B_1)> \frac{7}{4}\) gives a tighter lower bound \(B_1\) for \(C^*\).
On the other hand, taking \(B_2 < 2.0386\) with \(\Sigma(B_2)\le \frac{7}{4}\) gives a tighter upper bound \(B_2\) for \(C^*\). When \(B_1\) and \(B_2\) are sufficiently close, the inclusion \(C^*\in [B_1, B_2]\) implies that \(C^*\) has a very tight range of possible values.

At the end of this subsection, we discuss how to obtain a sufficiently tight interval \([B_1, B_2]\). Essentially, the problem reduces to how to compute \(\Sigma(B)\) exactly or how to give tight upper or lower bounds on \(\Sigma(B)\).
There are three points to note regarding the computation of \(\Sigma(B)\).

\begin{enumerate}
\item Since \(\Sigma(B) = \sum_{n=3}^\infty \lceil d_{n}(B) \rceil 2^{-n}\), where
\[
d_{n}(B)=\sup_{2\le k \in \mathbb{N}} \{nk- c_{B}(k)\}=\sup_{2\le k \in \mathbb{N}} \Big\{nk-\min \big\{1+(k-1)\big[1+(B-1)\sigma_k\big],Bk\log_2 k\big\} \Big\},
\]
to compute \(\Sigma(B)\) accurately and efficiently, we need a convenient method for computing \(d_{n}(B)\). Through analysis, we obtain the following expression.
\begin{equation} \label{eq15}
\begin{aligned}
	d_{n}(B)\! = \! \begin{cases}
		2n - 2B, & \text{if } n \le (1 + \frac{1}{\ln 2})B,  \\
		\max\big\{\widetilde{T}_n(\lfloor x_1^*\rfloor), \widetilde{T}_n(\lceil x_1^* \rceil)\big\}, & \text{if } (1\!+\!\frac{1}{\ln 2})B < n \le (B\!-\!1)Z_{0}\!+\!1, \\
		\max\big\{\widetilde{T}_n(\lfloor x_1^*\rfloor), \widetilde{T}_n(\lceil x_1^* \rceil), \widetilde{S}_n(\lfloor x_2^*\rfloor), \widetilde{S}_n(\lceil x_2^* \rceil)\big\}, & \text{if } n > (B-1)Z_{0}+1,
	\end{cases} \\
\end{aligned}
\end{equation}
where
\begin{equation*}
\begin{aligned}
	\widetilde{T}_n(x) &\triangleq nx - Bx \log_2 x, \quad x \ge 2,	\\
	\widetilde{S}_n(x) &\triangleq (n - 1)x - (x - 1)(B - 1)\sigma_x, \quad x \ge 2,\\
	x_1^* & \triangleq 2^{\frac{n}{B} - \frac{1}{\ln 2}},\\
	x_2^* & \triangleq 1 + 2^{y^*(1 - h(\frac{1}{y^*}))},\\
	Z_0   & \triangleq \min \left \{y+\dfrac{1}{\ln(2-\frac{2}{y})} \ \Big| \ y>2 \right\}.
\end{aligned}
\end{equation*}
Here, $\sigma_x$ satisfies Equation~\eqref{eq:sigma}, and \(y^*\) is the largest root of equation \(y + \frac{1}{\ln(2 - \frac{2}{y})} = \frac{n-1}{B-1}\). The detailed proof of Equation~\eqref{eq15} is given in Appendix~\ref{appexdix:eq15}.
\item To improve the lower bound \(2\) for \(C^*\), we seek \(B_1>2\) such that \(\Sigma(B_1)> \frac{7}{4}\).
By \(\Sigma(2.0386)\le \frac{7}{4}\) and Lemma~\ref{lemma9}, it follows that \(B_1< 2.0386\). Therefore, we obtain $c_{B_1}(2)=\min\{1+(2-1)(2B_1-1),2B_1\}=2B_1$, and hence
\[
\begin{aligned}
  d_n(B_1)&=\sup_{2\le k\in\mathbb{N}}\{nk-c_{B_1}(k)\}   \\
          &\ge 2n-c_{B_1}(2)  \\
          & =2(n-B_1)   \\
          & >0
\end{aligned}
\]
for all $3\le n\in \mathbb{N}$. Thus, it follows that
\[
  \Sigma(B_1)=\Sigma_N(B_1)+\sum_{n=N+1}^\infty \lceil d_{n}(B_1) \rceil 2^{-n}>\Sigma_N(B_1).
\]
Therefore, when explicitly computing \(\Sigma(B_1)\) in practice, it suffices to find \(N\) such that the \(N\)-th partial sum \(\Sigma_N(B_1)\) satisfies \(\Sigma_N(B_1)> \frac{7}{4}\); in this case, \(\Sigma(B_1)>\Sigma_N(B_1)> \frac{7}{4}\).
By numerical computation, we obtain that when \(B_1=
2.0001247570361019786170117629757988625444\), \(\Sigma(B_1)> \frac{7}{4}\).
Thus, the lower bound for \(C^*\) is improved from \(2\) to \(2.0001247570361019786170117629757988625444\).
The details of the computation are given in Appendix~\ref{lower}.
The complete computer simulation code for Appendix~\ref{lower} is publicly available in the repository~\cite{Zheng}; 
running it reproduces every certified bound reported in this paper.
\item To improve the upper bound \(2.0386\) for \(C^*\), we seek \(B_2<2.0386\) such that \(\Sigma(B_2)\le \frac{7}{4}\).
When computing \(\Sigma(B_2)\) exactly, we split the summation into the following two parts.
\[
  \Sigma(B_2)=\Sigma_L(B_2)+\sum_{n=L+1}^\infty \lceil d_{n}(B_2)\rceil 2^{-n}.
\]
Using Equation~\eqref{eq15}, we accurately compute the sum of the first \(L-2\) terms in $\Sigma_L(B_2)$.
We now analyze how to give an upper bound for the tail sum $\sum_{n=L+1}^\infty \lceil d_{n}(B_2)\rceil 2^{-n}$.

We obtain
\begin{equation} \label{eq16}
\begin{aligned}
  \widetilde{T}_n(x_1^*)&=nx_1^* - Bx_1^* \log_2 x_1^*    \\
                        &=\left[ n - B \left( \frac{n}{B} - \frac{1}{\ln 2} \right) \right] 2^{\frac{n}{B} - \frac{1}{\ln 2}} \\
                        &= \frac{B}{\mathrm{e}\ln 2} 2^{\frac{n}{B}}>0.
\end{aligned}
\end{equation}
Since \(x_2^*=1 + 2^{y^*(1 - h(\frac{1}{y^*}))}\), it follows that \(y^*\) satisfies equation \(h\left(\frac{1}{y}\right) + \frac{1}{y} \log_2(x_2^* - 1) = 1\). Therefore, \(y^*=\sigma_{x_{2^*}}\).
Moreover, because
\[
       n-1=(B-1)\Big(y^* + \frac{1}{\ln(2 - \frac{2}{y^*})} \Big),
\]
we further obtain
\begin{equation} \label{eq17}
	\begin{aligned}
		\widetilde{S}_n(x_2^*)&=(n-1)x_2^* - (x_2^*-1)(B-1)y^*  \\
		&=(B-1)\Big(y^* + \frac{1}{\ln(2 - \frac{2}{y^*})} \Big)x_2^* - (x_2^*-1)(B-1)y^*  \\
		&=(B-1)\Big(\frac{x_2^*}{\ln(2-\frac{2}{y^*})}+y^*\Big)>0.
	\end{aligned}
\end{equation}

From the differentiation and related analysis in Appendix~\ref{appexdix:eq15}, it follows that
\[
d_{n}(B)\le \max\big\{\widetilde{T}_n(x_1^*), \widetilde{S}_n( x_2^*)\big\}
\]
for all $n>(B-1)Z_{0}+1$. Therefore, when $L>(B_2-1)Z_{0}$, we obtain
\begin{equation} \label{eq18}
	\begin{aligned}
\sum_{n=L+1}^\infty\lceil d_{n}(B_2)\rceil 2^{-n} &= \sum_{n=L+1}^\infty\Big[ \lceil d_{n}(B_2) \rceil -d_n(B_2) \Big] 2^{-n}+\sum_{n=L+1}^\infty d_{n}(B_2) 2^{-n} \\
   &< \sum_{n=L+1}^\infty 2^{-n} + \sum_{n=L+1}^\infty \max\big\{\widetilde{T}_n(x_1^*), \widetilde{S}_n( x_2^*)\big\} 2^{-n}   \\
   & \overset{(a)}{<} 2^{-L}+\sum_{n=L+1}^\infty\widetilde{T}_n(x_1^*)2^{-n}+ \sum_{n=L+1}^\infty \widetilde{S}_n(x_2^*)2^{-n},
    \end{aligned}
\end{equation}
where $(a)$ follows from Equations~\eqref{eq16} and~\eqref{eq17}.
Equation~\eqref{eq16} further yields
\begin{equation} \label{eq19}
 \sum_{n=L+1}^\infty\widetilde{T}_n(x_1^*)2^{-n}=\sum_{n=L+1}^\infty\frac{B}{\mathrm{e}\ln 2} 2^{\frac{n}{B}-n}=\frac{B}{\mathrm{e}\ln 2} \cdot \frac{2^{(L+1)(\frac{1}{B}-1)}}{1-2^{\frac{1}{B}-1}}.
\end{equation}
Through analysis, we obtain the following inequality
\begin{equation} \label{eq20}
\begin{aligned}
\sum_{n=L+1}^\infty \widetilde{S}_n(x_2^*) 2^{-n} &< (L+1)2^{-L} + (B-1) \left( \frac{1}{\ln(2 - \frac{2}{3.4213})} - \frac{1}{\ln 2} \right) 2^{-L} \\
&\quad + \frac{B-1}{\ln(2 - \frac{2}{3.4213})} 2^{-1 - \frac{2}{\ln 2}} \cdot \frac{2^{\frac{1}{U\ln2}}}{U} \cdot \frac{2^{\beta(1-L)}}{2^\beta - 1},
\end{aligned}
\end{equation}
where $\beta \triangleq 1 - \frac{1}{B-1}$ and $U\triangleq \frac{L}{B-1} - \frac{1}{\ln 2}$. The detailed proof of Equation~\eqref{eq20} is given in Appendix~\ref{appexdix:eq20}.
From Equations~\eqref{eq18},~\eqref{eq19}, and~\eqref{eq20}, we obtain the following inequality, thereby obtaining an upper bound for the tail sum $\sum_{n=L+1}^\infty \lceil d_{n}(B_2)\rceil 2^{-n}$.
\begin{equation} \label{eq99}
\begin{aligned}
\sum_{n=L+1}^\infty\lceil d_{n}(B_2)\rceil 2^{-n} &< (L+2)2^{-L} + (B_2-1) \left( \frac{1}{\ln(2 - \frac{2}{3.4213})} - \frac{1}{\ln 2} \right) 2^{-L} \\
&\quad + \frac{B_2-1}{\ln(2 - \frac{2}{3.4213})} 2^{-1 - \frac{2}{\ln 2}} \cdot \frac{2^{\frac{1}{U\ln2}}}{U} \cdot \frac{2^{\beta(1-L)}}{2^\beta-1}+\frac{B_2}{\mathrm{e}\ln 2} \cdot \frac{2^{(L+1)(\frac{1}{B_2}-1)}}{1-2^{\frac{1}{B_2}-1}}.
\end{aligned}
\end{equation}
Therefore, if the exact sum of the first \(L-2\) terms $\Sigma_L(B_2)$  plus an upper bound on the tail sum is at most \(\frac{7}{4}\), then we can conclude that \(\Sigma(B_2)\le \frac{7}{4}\).
By numerical computation, we obtain that when \(B_2=
2.0001247570361019786170117629757988625463\), \(\Sigma(B_2)\le \frac{7}{4}\).
Thus, the upper bound for \(C^*\) is improved from \(2.0386\) to \(2.0001247570361019786170117629757988625463\).
The details of the computation are given in Appendix~\ref{lower}.
\end{enumerate}
In summary, the range of the optimal expansion factor \(C^*\) is narrowed from \([2, 2.0386]\) to \[ [2.0001247570361019786170117629757988625444, 2.0001247570361019786170117629757988625463]. \]

We discuss the optimal UCI corresponding to \(C^*\), considering the following two cases.
If there exists \(B_0 \in \{ B\ge1 \mid \Sigma(B)\le \frac{7}{4} \}\) such that \(B_0 = \inf \{ B\ge1 \mid \Sigma(B)\le \frac{7}{4} \}\), then \(C^* = B_0\). In this case, the code \(\mathcal{C}(B_0)\) obtained from Construction~\ref{con1} is optimal UCI.
Otherwise, if no such \(B_0\) exists, there is a sequence \(\{b_n\}_{n=1}^\infty\) with \(b_n \in \{ B\ge1 \mid \Sigma(B)\le \frac{7}{4} \}\) and \(\lim_{n\to\infty} b_n = \inf \{ B\ge1 \mid \Sigma(B)\le \frac{7}{4} \}\), then \(C^* = \lim_{n\to\infty} b_n\). In this case, the family of UCIs \(\{\mathcal{C}(b_n)\}_{n=1}^\infty\) obtained from Construction~\ref{con1} is optimal UCI. Therefore, in theory, the optimal UCI has already been found via Theorem~\ref{thm8}$(3)$ and Construction~\ref{con1}.
In practice, however, the UCI inequality and numerical computation show that the optimal expansion factor is $C^* = 2.00012475703610197861701176297579886254\cdots$. Thus, the optimal UCI is either \(\mathcal{C}(C^*)\), or, by taking a sequence \(\{b_n\}_{n=1}^\infty\) with \(\lim_{n\to\infty} b_n=C^*\) and \(b_n>C^*\) for all $n$, the corresponding code family \(\{\mathcal{C}(b_n)\}_{n=1}^\infty\) is optimal.

\section{Conclusions}\label{sec_con}
In this paper, we construct the one-point plus uniform-tail distributions and prove that they
are least favorable: for every UCI, a distribution in this family attains the worst-case ratio.
These distributions are the key ingredient in deriving the UCI inequality, which relates to UCI
as the Kraft inequality relates to prefix codes, and which compares any prescribed real number
$B$ with $C^{*}$. Using the UCI inequality, we obtain an equivalent definition of $C^{*}$; a
numerical computation based on it gives $C^{*}=2.00012475703610197861701176297579886254\cdots$, with the first $38$
decimal digits certified. Once $C^{*}$ is determined, an optimal UCI is obtained in theory.

The optimal UCI obtained in this way is, however, not described as concisely and clearly as the
classical $\gamma$ code proposed by Elias~\cite{Elias75}, nor as the $\nu$ code~\cite{Yan26},
which recently brought the upper bound on $C^{*}$ down to $2.0386$. It is therefore natural to
seek a construction of the optimal UCI whose codewords admit equally simple mathematical
expressions. Another direction is to ask whether the constant $2.00012475703610197861701176297579886254\cdots$ admits
a closed form.

\appendices
\section{Another Proof of Theorem~\ref{thm4}}\label{another}
For completeness we give a second, independent proof of Theorem~\ref{thm4}.
It is direct, but it requires the statement of the theorem to be guessed in advance; 
the proof in Section~\ref{new}, by contrast, follows the natural route through Theorem~\ref{thm3} and Corollary~\ref{cor2}.
\begin{proof}
	For every \(k\ge2\), the distributions \(P_{\varepsilon_k,k}\) and \(U_k\) belong to \(\mathcal{P}_\mathbb{N}\), and hence
	\[
	C_{\mathcal{C}}^{*}=\sup_{P\in\mathcal{P}_\mathbb{N}}\dfrac{A_{P}(L_\mathcal{C})}{\max\{1,H(P)\}}	\ge \sup_{P\in\{P_{\varepsilon_k,k},U_k \mid 2\leq k\in\mathbb{N}\}}\dfrac{A_{P}(L_\mathcal{C})}{\max\{1,H(P)\}}=T(\mathcal{C}).
	\]
	Therefore, it suffices to prove that
	\[
	\sup_{P\in\mathcal{P}_\mathbb{N}}\dfrac{A_{P}(L_\mathcal{C})}{\max\{1,H(P)\}}	\le T(\mathcal{C}),
	\]
	which is equivalent to proving that
	\begin{equation}  \label{eq10}
		A_{P}(L_\mathcal{C}) \le  T(\mathcal{C})\max\{1,H(P)\}
	\end{equation}
	for all $P\in\mathcal{P}_\mathbb{N}$.
	
	By Lemma~\ref{lemma2}, there exists a probability distribution $W=(w_1,w_2,\ldots)$ such that \(P =\sum_{k=1}^\infty w_k U_k\).		
	For  every \(k\ge2\), let
	\[
	\begin{aligned}
		\beta_k & \triangleq \frac{k\varepsilon_k}{k-1}\in(0,1],  \\
		R & \triangleq \sum_{k=2}^\infty \frac{w_k}{\beta_k}(1-\beta_k).
	\end{aligned}			
	\]
	We obtain
	\[
	\begin{aligned}
		\big[(1-\beta_k)U_1+\beta_kU_k\big](1)&=1-\beta_k+\frac{\beta_k}{k}=1-\varepsilon_k=P_{\varepsilon_k,k}(1),   \\
		\big[(1-\beta_k)U_1+\beta_kU_k\big](n)&=0+\frac{\beta_k}{k}=\frac{\varepsilon_k}{k-1}=P_{\varepsilon_k,k}(n)
	\end{aligned}
	\]
	for all $2\le n\le k$. Therefore, we have $P_{\varepsilon_k,k}=(1-\beta_k)U_1+\beta_kU_k$.
	We now consider two cases.
	\begin{enumerate}
		\item Case $R\le w_1$: Let $\lambda_1=w_1-R\ge 0$ and $\lambda_k=\frac{w_k}{\beta_k}$ for all $k\ge 2$.
		Then
		\[
		\begin{aligned}
			\sum_{k=1}^\infty \lambda_k =w_1-R&+\sum_{k=2}^\infty \frac{w_k}{\beta_k}=\sum_{k=1}^\infty w_k=1,  \\
			\lambda_1U_1+\sum_{k=2}^\infty \lambda_kP_{\varepsilon_k,k}&=\lambda_1U_1+\sum_{k=2}^\infty \lambda_k\big[(1-\beta_k)U_1+\beta_kU_k \big]   \\
			&=\Big[\lambda_1+\sum_{k=2}^\infty \frac{w_k}{\beta_k}(1-\beta_k)\Big]U_1+\sum_{k=2}^\infty w_kU_k   \\
			&= \sum_{k=1}^\infty w_kU_k   \\
			&= P.  \\
		\end{aligned}
		\]
		Furthermore, we obtain
		\[
		\begin{aligned}
			A_{P}(L_\mathcal{C})&= \sum_{n=1}^\infty  \Big[\lambda_1U_1+\sum_{k=2}^\infty \lambda_kP_{\varepsilon_k,k} \Big](n)L_\mathcal{C}(n)   \\
			&=\lambda_1 A_{U_1}(L_\mathcal{C})+\sum_{k=2}^\infty \lambda_k A_{P_{\varepsilon_k,k}}(L_\mathcal{C})  \\
			&\le \lambda_1 L_\mathcal{C}(1)+\sum_{k=2}^\infty \lambda_k T(\mathcal{C}) \\
			&\overset{(a)}{\le} \sum_{k=1}^\infty \lambda_k T(\mathcal{C})  \\
			& = T(\mathcal{C}),
		\end{aligned}
		\]
		where $(a)$ is due to
		\[
		T(\mathcal{C})\ge \frac{A_{U_2}(L_\mathcal{C})}{\max\{1,H(U_2)\}}=\frac{1}{2}\Big(L_\mathcal{C}(1)+L_\mathcal{C}(2)\Big)\ge L_\mathcal{C}(1).
		\]
		Therefore, Equation~\eqref{eq10} holds for all $P\in\mathcal{P}_\mathbb{N}$.
		\item Case $R> w_1$:
		We prove that the coefficients \(\{\lambda_k\}_{k=2}^\infty \) can be chosen such that
		\[
		0\le \lambda_k\le\frac{w_k}{\beta_k},
		\qquad
		\sum_{k=2}^\infty \lambda_k(1-\beta_k)=w_1.
		\]
		We now give a procedure for choosing the coefficients.
		Start with the initial remaining
		demand $r_1\triangleq w_1$, then process the indices in increasing order $k=2,3,\dots$.
		For each $k \ge 2$, proceed as follows.
		\begin{enumerate}
			\item If $r_{k-1}=0$, stop and set $\lambda_j\triangleq0$ for all $j\ge k$.
			\item If $\beta_k=1$, then $1-\beta_k=0$, so the choice of $\lambda_k$ is immaterial;
			set $\lambda_k\triangleq0$ and $r_k\triangleq r_{k-1}$.
			\item If $0<\beta_k<1$, set
			\[
			\lambda_k\triangleq\min\Bigl\{\frac{w_k}{\beta_k},\ \frac{r_{k-1}}{1-\beta_k}\Bigr\},
			\]
			and update the remaining demand to $r_k\triangleq r_{k-1}-\lambda_k(1-\beta_k)$.
		\end{enumerate}
		Because
		\[
		\sum_{k=2}^\infty \frac{w_k}{\beta_k}(1-\beta_k)=R>w_1=r_1,
		\]
		there exists a finite index \(K\) such that $\sum_{k=2}^K \frac{w_k}{\beta_k}(1-\beta_k)>w_1$.
		From the above selection process, it follows that the last nonzero coefficient equals $\frac{r_{k-1}}{1-\beta_k}$, and
		$$\sum_{k=2}^\infty \lambda_k(1-\beta_k)=\sum_{k=2}^K \lambda_k(1-\beta_k)=r_1=w_1.$$
		
		Define $t_k\triangleq w_k-\beta_k\lambda_k\ge0$ for all $k\ge2$. We obtain
		\[
		\begin{aligned}
			\sum_{k=2}^\infty \lambda_kP_{\varepsilon_k,k}+\sum_{k=2}^\infty t_kU_k &=
			\sum_{k=2}^\infty \lambda_k\big[(1-\beta_k)U_1+\beta_kU_k\big]+\sum_{k=2}^\infty (w_k-\beta_k\lambda_k)U_k  \\
			& =\sum_{k=2}^\infty \lambda_k(1-\beta_k)U_1+\sum_{k=2}^\infty w_kU_k  \\
			& =\sum_{k=1}^\infty w_kU_k   \\
			& =P,
		\end{aligned}
		\]
		and
		\[
		\begin{aligned}
			\sum_{k=2}^\infty(\lambda_k+t_k) &= \sum_{k=2}^\infty\lambda_k + \sum_{k=2}^\infty(w_k-\beta_k\lambda_k)\\
			&=\sum_{k=2}^\infty w_k+\sum_{k=2}^\infty \lambda_k(1-\beta_k)\\
			&=(1-w_1)+w_1\\
			&=1.
		\end{aligned}
		\]
		From the concavity of entropy, it follows that
		\[
		\begin{aligned}
			H(P) & = H\left(\sum_{k=2}^\infty \lambda_kP_{\varepsilon_k,k}+\sum_{k=2}^\infty t_kU_k\right)   \\
			& \ge \sum_{k=2}^\infty \lambda_kH(P_{\varepsilon_k,k}) +\sum_{k=2}^\infty t_kH(U_k) \\
			& = \sum_{k=2}^\infty \lambda_k +\sum_{k=2}^\infty t_k\log_2k.
		\end{aligned}
		\]
		By the definition of \(T(\mathcal{C})\), we have
		\[
		\begin{aligned}
			A_{P}(L_\mathcal{C})&= \sum_{n=1}^\infty  \Big[\sum_{k=2}^\infty \lambda_kP_{\varepsilon_k,k}+\sum_{k=2}^\infty t_kU_k\Big](n)L_\mathcal{C}(n)   \\
			&=\sum_{k=2}^\infty \lambda_k A_{P_{\varepsilon_k,k}}(L_\mathcal{C})+\sum_{k=2}^\infty t_k A_{U_k}(L_\mathcal{C}) \\
			&\le \sum_{k=2}^\infty \lambda_k T(\mathcal{C})+ \sum_{k=2}^\infty t_k H(U_k)T(\mathcal{C})  \\
			&= T(\mathcal{C})\Big[\sum_{k=2}^\infty \lambda_k +\sum_{k=2}^\infty t_k\log_2k\Big]   \\
			& \le T(\mathcal{C})H(P).
		\end{aligned}
		\]
		Therefore, Equation~\eqref{eq10} holds for all $P\in\mathcal{P}_\mathbb{N}$.
	\end{enumerate}
	This completes the proof.
\end{proof}

\section{Proof of Equation~\eqref{eq15}} \label{appexdix:eq15}
We first express \(d_n(B)\) in terms of \(\widetilde{S}_n(x)\) and \(\widetilde{T}_n(x)\). We obtain
\[
\begin{aligned}
d_n(B) &=\sup_{2\le k \in \mathbb{N}} \Big\{nk-\min \big\{1+(k-1)\big[1+(B-1)\sigma_k\big],Bk\log_2 k\big\} \Big\}   \\
   & = \sup_{2\le k \in \mathbb{N}} \max\big\{(n-1)k-(k-1)(B-1)\sigma_k,nk-Bk\log_2 k \big\}   \\
   &= \sup_{2\le k \in \mathbb{N}} \max \big\{\widetilde{S}_n(k), \widetilde{T}_n(k) \big\}.
\end{aligned}
\]
Second, we prove that \(\sup\) and \(\max\) can be interchanged; that is,
\begin{equation} \label{eq21}
\begin{aligned}
	d_n(B) &= \sup_{2\le k \in \mathbb{N}} \max \big\{\widetilde{S}_n(k), \widetilde{T}_n(k) \big\}  \\
	& = \max \left\{\sup_{2\le k \in \mathbb{N}}\widetilde{S}_n(k), \sup_{2\le k \in \mathbb{N}}\widetilde{T}_n(k)  \right\}.
\end{aligned}
\end{equation}
\begin{enumerate}
\item Since \(\widetilde{S}_n(k)\le\sup_{2\le k \in \mathbb{N}}\widetilde{S}_n(k)\) and \(\widetilde{T}_n(k)\le\sup_{2\le k \in \mathbb{N}}\widetilde{T}_n(k)\), we have
\[
\max \big\{\widetilde{S}_n(k), \widetilde{T}_n(k) \big\} \le \max \left\{\sup_{2\le k \in \mathbb{N}}\widetilde{S}_n(k), \sup_{2\le k \in \mathbb{N}}\widetilde{T}_n(k)  \right\},
\]
and hence
\begin{equation} \label{eq22}
\sup_{2\le k \in \mathbb{N}} \max \big\{\widetilde{S}_n(k), \widetilde{T}_n(k) \big\}\le \max \left\{\sup_{2\le k \in \mathbb{N}}\widetilde{S}_n(k), \sup_{2\le k \in \mathbb{N}}\widetilde{T}_n(k)  \right\}.
\end{equation}
\item Since \(\widetilde{S}_n(k)\le \max \big\{\widetilde{S}_n(k), \widetilde{T}_n(k) \big\}\), it follows that \(\sup_{2\le k \in \mathbb{N}}\widetilde{S}_n(k)\le \sup_{2\le k \in \mathbb{N}} \max \big\{\widetilde{S}_n(k), \widetilde{T}_n(k) \big\}\).
Similarly, we obtain \(\sup_{2\le k \in \mathbb{N}}\widetilde{T}_n(k)\le \sup_{2\le k \in \mathbb{N}} \max \big\{\widetilde{S}_n(k), \widetilde{T}_n(k) \big\}\), and hence
\begin{equation} \label{eq23}
\max \left\{\sup_{2\le k \in \mathbb{N}}\widetilde{S}_n(k), \sup_{2\le k \in \mathbb{N}}\widetilde{T}_n(k)  \right\}\le \sup_{2\le k \in \mathbb{N}} \max \big\{\widetilde{S}_n(k), \widetilde{T}_n(k) \big\}.
\end{equation}
\end{enumerate}
By Equations~\eqref{eq22} and~\eqref{eq23}, Equation~\eqref{eq21} holds. From Equation~\eqref{eq21}, the computation of \(d_n(B)\) reduces to that of \(\sup_{2\le k \in \mathbb{N}}\widetilde{S}_n(k)\) and \(\sup_{2\le k \in \mathbb{N}}\widetilde{T}_n(k)\).

Next, we compute \(\sup_{2\le k \in \mathbb{N}}\widetilde{S}_n(k)\). The idea of the computation is to differentiate \(\widetilde{S}_n(x)\) and thereby determine its monotonicity over \([2,\infty)\).
Since \(\widetilde{S}_n(x)= (n-1)x-(x-1)(B-1)\sigma_x\) is implicitly defined as a function of \(x\), it cannot be differentiated directly. Let \(y=\sigma_x\ge 2\), we obtain
\begin{equation}  \label{eq24}
 h\left(\frac{1}{y}\right)+\frac{1}{y}\log_{2}(x-1)=1.
\end{equation}
Multiplying Equation~\eqref{eq24} by \(\ln 2\) and simplifying, we obtain
\[
 \ln\frac{y}{y-1}+\frac{1}{y}\ln\left[(x-1)(y-1)\right]-\ln 2=0.
\]
Taking the derivative yields
\[
 y'=\dfrac{y}{(x-1)\ln\left[(x-1)(y-1)\right]},
\]
and hence
\[
\begin{aligned}
 \widetilde{S}_n'(x) & =n-1-(B-1)y-y'(x-1)(B-1)    \\
                     & =n-1-(B-1)\left[y+\frac{1}{\ln(2-\frac{2}{y})}\right]. \\
\end{aligned}
\]
Let $Z(y)\triangleq y+\frac{1}{\ln(2-\frac{2}{y})}$ for $y\in(2,\infty)$.
Differentiating shows that \(Z(y)\) is first decreasing and then increasing over \((2,\infty)\). Let \(Z_0 \triangleq \min \Big\{y+\dfrac{1}{\ln(2-\frac{2}{y})} \ \Big| \ y>2 \Big\}\). Then \(Z_0\) is the minimum value of \(Z(y)\) over \((2,\infty)\).
\begin{enumerate}
\item Case $n-1\le(B-1)Z_0$: In this case, we obtain
\[
 \widetilde{S}_n'(x)=n-1-(B-1)Z(y)\le n-1-(B-1)Z_0\le 0.
\]
Therefore, $\widetilde{S}_n(x)$ is nonincreasing on the interval \([2,\infty)\), and hence we obtain $\widetilde{S}_n(x)\le \widetilde{S}_n(2)$.
Thus,
\begin{equation}  \label{eq25}
\sup_{2\le k \in \mathbb{N}}\widetilde{S}_n(k)=\widetilde{S}_n(2)=2n-2B.
\end{equation}
\item Case $n-1>(B-1)Z_0$: From Equation~\eqref{eq24}, we have \(x=1+2^{y-yh(\frac{1}{y})}\). Differentiating, we obtain that \(x\) is strictly increasing in \(y\) on the interval \([2,\infty)\).
Let $\widetilde{S}_n'(x)=0$, then
$$Z(y)=\frac{n-1}{B-1}>Z_0.$$
Since \(Z(y)\) is first decreasing and then increasing over \((2,\infty)\), the  equation $Z(y)=\frac{n-1}{B-1}$ has two roots, denoted by \(y_1\) and \(y_2\), with \(y_1<y_2\). Therefore, the solution to $\widetilde{S}_n'(x)<0$ is \(2<y<y_1\) or \(y>y_2\); the solution to $\widetilde{S}_n'(x)>0$ is \(y_1<y<y_2\).
Since \(x\) is strictly increasing in \(y\) on the interval \([2,\infty)\), it follows that $\widetilde{S}_n(x)$ is decreasing on \([2,1+2^{y_1-y_1h(\frac{1}{y_1})}]\), increasing on \([1+2^{y_1-y_1h(\frac{1}{y_1})},1+2^{y_2-y_2h(\frac{1}{y_2})}]\), and decreasing on \([1+2^{y_2-y_2h(\frac{1}{y_2})},\infty)\), and hence
\[
  \widetilde{S}_n(x)\le \max\{\widetilde{S}_n(2),\widetilde{S}_n(x_2^*)\},
\]
where $x_2^*=1+2^{y_2-y_2h(\frac{1}{y_2})}$. Thus,
\begin{equation}  \label{eq26}
	\sup_{2\le k \in \mathbb{N}}\widetilde{S}_n(k)=\max\big\{\widetilde{S}_n(2)=2n\!-\!2B,\widetilde{S}_n(\lfloor x_2^* \rfloor),\widetilde{S}_n(\lceil x_2^* \rceil)\big\}.
\end{equation}
\end{enumerate}

Finally, we compute \(\sup_{2\le k \in \mathbb{N}}\widetilde{T}_n(k)\).
The idea of the computation is to differentiate \(\widetilde{T}_n(x)\) and thereby determine its monotonicity over \([2,\infty)\).
By taking the derivative, we obtain
\[
 \widetilde{T}_n'(x)=n-B\left(\log_{2}x+\frac{1}{\ln2}\right).
\]
Therefore, $\widetilde{T}_n(x)$ is increasing on \((0,x_1^*]\), and decreasing on \([x_1^*,\infty)\), where $x_1^*=2^{\frac{n}{B}-\frac{1}{\ln 2}}$.
We now consider two cases depending on whether \(x_1^*\le2\) or \(x_1^*>2\).
\begin{enumerate}
\item Case \(x_1^*\le2\): In this case, $\frac{n}{B}-\frac{1}{\ln 2}\le 1$, which is equivalent to $n\le(1+\frac{1}{\ln 2})B$. The function $\widetilde{T}_n(x)$ is decreasing on \([2,\infty)\). Therefore, $\widetilde{T}_n(x)\le \widetilde{T}_n(2)$, and hence
\begin{equation}  \label{eq27}
	\sup_{2\le k \in \mathbb{N}}\widetilde{T}_n(k)=\widetilde{T}_n(2)=2n-2B.
\end{equation}
\item Case \(x_1^*>2\): In this case, $n>(1+\frac{1}{\ln 2})B$ and $\widetilde{T}_n(x)$ is increasing on \([2,x_1^*]\), and decreasing on \([x_1^*,\infty)\). Therefore, $\widetilde{T}_n(x)\le \widetilde{T}_n(x_1^*)$, and hence
\begin{equation}  \label{eq28}
	\sup_{2\le k \in \mathbb{N}}\widetilde{T}_n(k)=\max\big\{\widetilde{T}_n(\lfloor x_1^* \rfloor),\widetilde{T}_n(\lceil x_1^* \rceil)\big\}\ge\widetilde{T}_n(2)=2n-2B .
\end{equation}
\end{enumerate}
In summary, from Equations~\eqref{eq21},~\eqref{eq25},~\eqref{eq26},~\eqref{eq27}, and~\eqref{eq28}, we obtain
\begin{equation*}
	\begin{aligned}
		d_{n}(B)\! = \! \begin{cases}
			2n - 2B, & \text{if } n \le (1 + \frac{1}{\ln 2})B,  \\
			\max\big\{\widetilde{T}_n(\lfloor x_1^*\rfloor), \widetilde{T}_n(\lceil x_1^* \rceil)\big\}, & \text{if } (1\!+\!\frac{1}{\ln 2})B < n \le (B\!-\!1)Z_{0}\!+\!1, \\
			\max\big\{\widetilde{T}_n(\lfloor x_1^*\rfloor), \widetilde{T}_n(\lceil x_1^* \rceil), \widetilde{S}_n(\lfloor x_2^*\rfloor), \widetilde{S}_n(\lceil x_2^* \rceil)\big\}, & \text{if } n > (B-1)Z_{0}+1.
		\end{cases} \\
	\end{aligned}
\end{equation*}
This completes the proof of Equation~\eqref{eq15}.

\section{Upper and lower bounds for $C^*$} \label{lower}
Since all numerical computations are performed over the reals, we carried them out in the ball arithmetic of FLINT~\cite{flint} to keep our conclusions rigorous.
A ball $[t\pm r]$ denotes the set $\{x \ \big| \ |x-t|\le r\}$ of real numbers within distance $r$ of $t$.
Ball arithmetic provides certified enclosures for the four basic arithmetic operations on real numbers.
Specifically, suppose we apply one of the four operations to two real numbers $x_1$ and $x_2$.
Let  $\oplus$ be a binary operation with domain $(\mathbb{R},\mathbb{R})$,
and let $\widetilde{\oplus}$ denote the operation in FLINT ball arithmetic corresponding to operation $\oplus$.
If
\[
x_1 \in [m_1 \pm r_1],
\qquad
x_2 \in [m_2 \pm r_2],
\]
then, for an operation $\oplus$,
\[
x_1\oplus x_2 \in [m_3\pm r_3],
\]
where $[m_3\pm r_3]=[m_1 \pm r_1]\widetilde{\oplus}[m_2 \pm r_2]$.
Note that $[m_3\pm r_3]$ may be \text{NaN} or an unbounded ball, such as $[0\pm\infty]$; this can occur, for example, when applying $\log$ to a ball containing zero.
We refer interested readers to the FLINT documentation~\cite{flint} for further details.

Based on Theorem~\ref{thm8}$(1)$ and $(2)$, we design the following algorithm, whose purpose is to obtain tighter upper and lower bounds on \(C^*\).
Let $\lb{s}$ and $\ub{s}$ denote the lower and upper endpoint of the ball $s$,
respectively; that is, if $s=[m\pm r]$, then $\lb{s}=m-r$ and $\ub{s}=m+r$.
The algorithm's core idea is as follows.
For each midpoint $B_m$, we first compute a ball $s_{\mathrm{trunc}}$ enclosing the sum of the first \(L-2\) terms $\Sigma_L(B_m)$.
If $\lb{s_{\mathrm{trunc}}}>\frac{7}{4}$, then $\Sigma(B_m)>\frac{7}{4}$, and hence $B_m$ is a lower bound for $C^*$.
Otherwise, we compute a ball $s_{\mathrm{tail}}$ enclosing an upper bound in Equation~\eqref{eq99} for the tail sum. If $\ub{s_{\mathrm{trunc}}}+\ub{s_{\mathrm{tail}}}\le\frac{7}{4}$, then $\Sigma(B_m)\le\frac{7}{4}$, and hence $B_m$ is an upper bound for $C^*$.
If neither test succeeds, double $L$ and repeat the tests.
The detailed steps are shown in Algorithm~\ref{alg:solve}.
We highlight several implementation details below.

\begin{algorithm}[t]
	\LinesNumbered
	\caption{Tighter upper and lower bounds on \(C^*\)}
	\label{alg:solve}
	\KwIn{Target precision $\epsilon$ in bits}
	\KwOut{Exact rational endpoints \(B_1<B_2\) such that
		$\Sigma(B_1)> \frac{7}{4}$ and
		$\Sigma(B_2)\le\frac{7}{4}$.}
	
	$B_1 \gets 2,\ B_2 \gets 2.0386\;(=10193/5000)$
	\tcp*[r]{initial values from~\cite{Yan26}}
	\nllabel{line:solve:init-bounds}
	
	\While{$B_2-B_1>2^{-\epsilon}$}{\nllabel{line:solve:outer-loop}
		$B_m\gets \frac{B_1+B_2}{2},\ L\gets 16$
		\nllabel{line:solve:midpoint}\;
		\While{\textbf{True}}{\nllabel{line:solve:inner-loop}
			$s_{\mathrm{trunc}}\gets
			\sum_{\ell=3}^{L}\lceil d_{\ell}(B_m)\rceil 2^{-\ell}$
			\tcp*[r]{with $(\epsilon+64)$-bit precision}\nllabel{line:solve:truncated-sum}
			\If{$\lb{s_{\mathrm{trunc}}}>\frac{7}{4}$}{\nllabel{line:solve:lower-test}
				$B_1\gets B_m$
				\nllabel{line:solve:lower-update}\;
				\textbf{break}
				\nllabel{line:solve:lower-break}
			}
			$s_{\mathrm{tail}}\gets \mathsf{TailSumUB}(B_m,L)$
			by Equation~\eqref{eq99}
			\tcp*[r]{with $(\epsilon+64)$-bit precision}
			\nllabel{line:solve:tail-bound}
			\If{$\ub{s_{\mathrm{trunc}}}+\ub{s_{\mathrm{tail}}}\le \frac{7}{4}$}
			{\nllabel{line:solve:upper-test}
				$B_2\gets B_m$
				\nllabel{line:solve:upper-update}\;
				\textbf{break}
				\nllabel{line:solve:upper-break}
			}
			$L\gets 2L$\nllabel{line:solve:double-truncation}
		}
	}
	\Return $B_1, B_2$\nllabel{line:solve:return}
\end{algorithm}

\begin{enumerate}
       \item The algorithm initializes the two bounds to \(2\) and \(2.0386\)~(\cref{line:solve:init-bounds}), values taken from~\cite{Yan26}. 
        At each iteration, the algorithm computes the midpoint ${B}_m$ of the upper and lower bounds. 
      By comparing the endpoints of the enclosure of \(\Sigma(B_m)\) with \(\frac74\), it determines whether \(B_m\) is an upper or lower bound, and thereby updates the bounds on \(C^*\). This uses bisection to approximate \(C^*\).
       The initial upper bound $B_2=2.0386$ satisfies $\Sigma(B_2)\le\frac74$ by Theorems~\ref{thm9} and~\ref{thm5} (the $\nu$ code), so it needs no numerical certificate; every later value of $B_2$ is certified by the upper-bound test.
\item Note that \(B_1\), \(B_2\), and \(B_m\) are all rational numbers, so the
bisection itself is carried out in exact rational arithmetic, without any
rounding. Ball arithmetic is used only to evaluate \(\Sigma(B_m)\), in which case
$s_{\mathrm{trunc}}$ (\cref{line:solve:truncated-sum}) and
$s_{\mathrm{tail}}$ (\cref{line:solve:tail-bound}) are returned as balls
(enclosures).
	The lower-bound test (\cref{line:solve:lower-test}) explicitly requires $\lb{s_{\mathrm{trunc}}}>\frac74$, and the upper-bound test (\cref{line:solve:upper-test}) requires $\ub{s_{\mathrm{trunc}}}+\ub{s_{\mathrm{tail}}}\le\frac74$.
	Thus, each update is justified by the appropriate endpoints of the ball enclosures. 
	The implementation actually uses the strict test
	$\ub{s_{\mathrm{trunc}}}+\ub{s_{\mathrm{tail}}}<\frac74$,
	which is slightly stronger than the non-strict test $\le\frac74$ shown in
	Algorithm~\ref{alg:solve}; all upper-bound certificates recorded in the logs satisfy both.
\item The algorithm takes a target precision $\epsilon$ (in bits) as input and evaluates $s_{\mathrm{trunc}}$ and $s_{\mathrm{tail}}$ with a fixed $(\epsilon+64)$-bit precision. The outer loop continues until the gap between the bounds is at most $2^{-\epsilon}$ (\cref{line:solve:outer-loop}).
	The inner loop repeats until either test succeeds (\cref{line:solve:inner-loop}). If neither test succeeds, the algorithm doubles \(L\) (\cref{line:solve:double-truncation}) and repeats the tests at the same computational precision, without imposing a preset upper limit on \(L\); in the implementation, the partial sum is extended by adding only the newly included terms $\ell=L+1,\dots,2L$, which is equivalent to recomputing the whole sum from $\ell=3$.
	Once the bounds are within the target gap, it returns them (\cref{line:solve:return}).
\item 
The ceiling $\lceil d_{\ell}(B_m)\rceil$ is not computed as a single integer;
it is kept as an interval enclosure, namely a ball that contains the integer
interval obtained by rounding up the two endpoints of the ball for
$d_{\ell}(B_m)$.  Since $x\mapsto\lceil x\rceil$ is nondecreasing, this interval
contains $\lceil d_{\ell}(B_m)\rceil$ for every value of $d_{\ell}(B_m)$
admitted by that ball.  It is a singleton precisely when the two rounded
endpoints coincide, in particular whenever the ball contains no integer.  All
subsequent operations are again ball operations, so $s_{\mathrm{trunc}}$ is a
rigorous enclosure of $\Sigma_L(B_m)$; in the runs below, these enclosures are
narrow enough for both tests to succeed at every iteration.
\end{enumerate}

We ran Algorithm~\ref{alg:solve} for target precisions \(\epsilon = 16, 32, 48, 64, 80, 96, 112,\) and \(128\). Table~\ref{tab2} lists the final lower bounds $B_1$, their truncation parameters $N$, and the positive offsets $\lb{s_{\mathrm{trunc}}}-\frac74$. Table~\ref{tab:certificates} lists the final upper bounds $B_2$, their truncation parameters $L$, and the negative offsets $\ub{s_{\mathrm{trunc}}}+\ub{s_{\mathrm{tail}}}-\frac74$.
The certificates in Tables~\ref{tab2} and~\ref{tab:certificates} come, respectively, from the last lower-bound and the last upper-bound update of each run, and are evaluated at the exact rational midpoints before any decimal rounding.
The program prints \(N\) on lower-update lines and \(L\) on upper-update lines; Table~\ref{tab2} reports \(N\) and Table~\ref{tab:certificates} reports \(L\).
The upper certificates are computed by adding the logged upper endpoints $\ub{s_{\mathrm{trunc}}}$ and $\ub{s_{\mathrm{tail}}}$, as in the algorithm's upper-bound test.
The columns $\lb{s_{\mathrm{trunc}}}-\frac74$ and $\ub{s_{\mathrm{trunc}}}+\ub{s_{\mathrm{tail}}}-\frac74$ report signed offsets from the threshold $\frac74$. The lower offsets in Table~\ref{tab2} are rounded down and the upper offsets in Table~\ref{tab:certificates} are rounded up, both to 12 significant digits, so the displayed values retain their respective lower- and upper-bound directions.

In the tables, for precisions \(\epsilon = 16\) and \(32\), we retain the exact values; for the remaining precisions, we round the bounds \(B_1\) and \(B_2\) down and up, respectively, to 40 decimal places. For example, when \(\epsilon = 48\), the output lower bound is \(B_1 = 2.0001247570361014140871702693402767181396484375\); retaining 40 decimal places requires truncation (rounding down) to obtain \(2.0001247570361014140871702693402767181396\). The output upper bound is \(B_2 = 2.00012475703610360824313829652965068817138671875\); retaining 40 decimal places requires rounding up to obtain \(2.0001247570361036082431382965296506881714\). As can be seen from Tables~\ref{tab2} and~\ref{tab:certificates}, we obtain the range that \(C^*\) must satisfy:
\[
2.0001247570361019786170117629757988625444 \le C^* \le 2.0001247570361019786170117629757988625463.
\]
Therefore, based on Tables~\ref{tab2} and~\ref{tab:certificates}, we can determine $38$ decimal places of \(C^*\), yielding
\[
C^* = 2.00012475703610197861701176297579886254\cdots.
\]
\begin{table}[t]
	\caption{Final lower bounds and their certificates for target precisions up to 128 bits.}
	\label{tab2}
	\centering
	\small
	\setlength{\tabcolsep}{4pt}
	\renewcommand{\arraystretch}{1.25}
	\begin{tabular}{crrc}
		\hline
		\(\epsilon\) & \(B_1\) & \(N\) & \(\lb{s_{\mathrm{trunc}}}-\frac74\) \\
		\hline
		16 & 2.000122509765625 & 32768 & $8.73602458947\times10^{-5}$ \\
		32 & 2.00012475701272487640380859375 & 262144 & $1.82280655647\times10^{-8}$ \\
		48 & 2.0001247570361014140871702693402767181396 & 524288 & $4.39569440980\times10^{-13}$ \\
		64 & 2.0001247570361019785962000683277040025132 & 524288 & $1.61870559115\times10^{-17}$ \\
		80 & 2.0001247570361019786170117578389928076135 & 1048576 & $3.98930252040\times10^{-24}$ \\
		96 & 2.0001247570361019786170117629682396196525 & 1048576 & $5.86692336423\times10^{-27}$ \\
		112 & 2.0001247570361019786170117629757988398337 & 1048576 & $1.75918126383\times10^{-32}$ \\
		128 & 2.0001247570361019786170117629757988625444 & 1048576 & $8.21130463268\times10^{-37}$ \\
		\hline
	\end{tabular}
\end{table}
\begin{table}[t]
	\caption{Final upper bounds and their certificates for target precisions up to 128 bits.}
	\label{tab:certificates}
	\centering
	\small
	\setlength{\tabcolsep}{4pt}
	\renewcommand{\arraystretch}{1.25}
	\begin{tabular}{crrc}
		\hline
		\(\epsilon\) & \(B_2\) & \(L\) & \(\ub{s_{\mathrm{trunc}}}+\ub{s_{\mathrm{tail}}}-\frac74\) \\
		\hline
		16 & 2.00013193359375 & 32768 & $-3.47932359547\times10^{-3}$ \\
		32 & 2.0001247571565210819244384765625 & 262144 & $-9.40390253595\times10^{-8}$ \\
		48 & 2.0001247570361036082431382965296506881714 & 262144 & $-6.30414300326\times10^{-13}$ \\
		64 & 2.0001247570361019786296802314140563794354 & 524288 & $-9.85366357591\times10^{-18}$ \\
		80 & 2.0001247570361019786170122687057391203635 & 524288 & $-3.47849709031\times10^{-22}$ \\
		96 & 2.0001247570361019786170117629760348275736 & 1048576 & $-1.83207969992\times10^{-28}$ \\
		112 & 2.0001247570361019786170117629757989587793 & 1048576 & $-7.45994450227\times10^{-32}$ \\
		128 & 2.0001247570361019786170117629757988625463 & 1048576 & $-5.82921078712\times10^{-37}$ \\
		\hline
	\end{tabular}
\end{table}

\section{Proof of Equation~\eqref{eq20}} \label{appexdix:eq20}
From Equation~\eqref{eq17}, it is known that
\begin{equation*}
\widetilde{S}_n(x_2^*)=(B-1)\Big(\frac{x_2^*}{\ln(2-\frac{2}{y^*})}+y^*\Big).
\end{equation*}
Therefore, we first bound the three terms \(\ln(2-\frac{2}{y^*})\), \(y^*\), and \(x_2^*\) using inequalities.
From Appendix~\ref{appexdix:eq15}, it is known that \(Z(y)=y + \frac{1}{\ln(2 - \frac{2}{y})}\) first decreases and then increases over \((2,\infty)\). Without loss of generality, we may assume that it is decreasing on \((2, y_{\min}]\) and increasing on \([y_{\min}, \infty)\).
By differentiation, we find that \(y_{\min} > 3.4213\). Because \(y^*\) is the largest root of equation \(y + \frac{1}{\ln(2 - \frac{2}{y})} = \frac{n-1}{B-1}\), we obtain $y^*>y_{\min} > 3.4213$, which further yields
\[
  \ln\Big(2-\frac{2}{y^*}\Big) > \ln\Big(2-\frac{2}{3.4213}\Big)>0.
\]
Let $y_0\triangleq \frac{n-1}{B-1}-\frac{1}{\ln2}$. Since $0<\ln(2-\frac{2}{y^*})<\ln 2$, we obtain
\[
  y^*=\frac{n-1}{B-1}-\frac{1}{\ln(2-\frac{2}{y^*})} < \frac{n-1}{B-1}-\frac{1}{\ln2} =y_0.
\]
Because $\Phi(y)=y\left(1-h\left(\frac{1}{y}\right)\right)$ is strictly increasing on \([2,\infty)\), and \(x_2^* =1 + 2^{y^*(1 - h(\frac{1}{y^*}))}\), we obtain
\[
  x_2^*-1=2^{\Phi(y^*)}<2^{\Phi(y_0)}.
\]

Second, substituting the above inequality into the sum term \(\sum_{n=L+1}^\infty \widetilde{S}_n(x_2^*) 2^{-n}\), we obtain
\begin{equation} \label{eq29}
	\begin{aligned}
\sum_{n=L+1}^\infty \widetilde{S}_n(x_2^*) 2^{-n} & = \sum_{n=L+1}^\infty (B-1)\Big(y^*+\frac{x_2^*}{\ln(2-\frac{2}{y^*})}\Big) 2^{-n}   \\
          & < \sum_{n=L+1}^\infty\!(B-1)\Big(y_0+\frac{1}{ \ln\big(2-\frac{2}{3.4213}\big)}\Big)2^{-n}\!+\!\sum_{n=L+1}^\infty\!(B-1)\cdot\frac{2^{\Phi(y_0)-n}}{\ln\big(2-\frac{2}{3.4213}\big)}  \\
          &= S_1+S_2,
   	\end{aligned}
\end{equation}
where
\begin{equation} \label{eq30}
	\begin{aligned}
  S_1& =\sum_{n=L+1}^\infty\!(B-1)\Big(y^*+\frac{x_2^*}{\ln(2-\frac{2}{y^*})}\Big)2^{-n}\\
  &=\sum_{n=L+1}^\infty(n-1)2^{-n} +(B-1)\left(\frac{1}{\ln\big(2-\frac{2}{3.4213}\big)}-\frac{1}{\ln2}\right)\sum_{n=L+1}^\infty 2^{-n}   \\
  & = (L+1)2^{-L}+(B-1)\left(\frac{1}{\ln\big(2-\frac{2}{3.4213}\big)}-\frac{1}{\ln2}\right)2^{-L}
   	\end{aligned}
\end{equation}
and
\begin{equation*}
	S_2 =\sum_{n=L+1}^\infty\!(B-1)\cdot\frac{2^{\Phi(y_0)-n}}{\ln\big(2-\frac{2}{3.4213}\big)}
	= \frac{B-1}{\ln\big(2-\frac{2}{3.4213}\big)}
	\sum_{n=L+1}^\infty2^{\Phi(y_0)-n}.
\end{equation*}

Next, we give an upper bound for $\sum_{n=L+1}^\infty2^{\Phi(y_0)-n}$.
Since $-\ln(1-x)\ge x$  holds for all $x<1$, we obtain
\[
\begin{aligned}
h\Big(\frac{1}{y_0}\Big)&=-\frac{1}{y_0}\log_{2}\frac{1}{y_0}-\Big(1-\frac{1}{y_0}\Big)\log_{2}\Big(1-\frac{1}{y_0}\Big)\\
 & \ge -\frac{1}{y_0}\log_{2}\frac{1}{y_0}+\Big(1-\frac{1}{y_0}\Big)\cdot \frac{1}{y_0\ln 2},
\end{aligned}
\]
and hence
\[
\begin{aligned}
2^{\Phi(y_0)}&=2^{y_0(1-h(\frac{1}{y_0}))}\\
	& \le 2^{y_0-\log_2{y_0}-\frac{1}{\ln 2}+\frac{1}{y_0\ln 2}}  \\
	& = \frac{2^{y_0}}{y_0}\cdot 2^{-\frac{1}{\ln 2}}\cdot 2^{\frac{1}{y_0\ln 2}}.
\end{aligned}
\]
Let $\beta \triangleq 1 - \frac{1}{B-1}$, then
\[
y_0-n=(n-1)\left(\frac{1}{B-1}-1\right)-1-\frac{1}{\ln 2}=-(n-1)\beta-1-\frac{1}{\ln 2}.
\]
Let $U\triangleq \frac{L}{B-1} - \frac{1}{\ln 2}$, then $U\ge\frac{3}{2.0386-1}-\frac{1}{\ln 2}>0$. If $n\ge L+1$, then $y_0=\frac{n-1}{B-1}-\frac{1}{\ln2} \ge U$, and hence
\[
 \frac{1}{y_0}\le  \frac{1}{U}, \qquad 2^{\frac{1}{y_0\ln 2}} \le 2^{\frac{1}{U\ln 2}}.
\]
Therefore, we obtain
\begin{equation} \label{eq31}
\begin{aligned}
\sum_{n=L+1}^\infty2^{\Phi(y_0)-n}& \le \sum_{n=L+1}^\infty\frac{2^{y_0-n}}{y_0}\cdot 2^{-\frac{1}{\ln 2}}\cdot 2^{\frac{1}{y_0\ln 2}}  \\
 &\le 2^{-\frac{1}{\ln 2}}\cdot\frac{2^{\frac{1}{U\ln 2}}}{U} \sum_{n=L+1}^\infty 2^{-(n-1)\beta-1-\frac{1}{\ln 2}} \\
 & =2^{-1 - \frac{2}{\ln 2}} \cdot \frac{2^{\frac{1}{U\ln2}}}{U} \cdot \frac{2^{\beta(1-L)}}{2^\beta - 1}.
\end{aligned}
\end{equation}

Finally, combining Equations~\eqref{eq29}, \eqref{eq30}, and \eqref{eq31}, we obtain
\begin{equation*}
\begin{aligned}
 	\sum_{n=L+1}^\infty \widetilde{S}_n(x_2^*) 2^{-n} &< (L+1)2^{-L} + (B-1) \left( \frac{1}{\ln(2 - \frac{2}{3.4213})} - \frac{1}{\ln 2} \right) 2^{-L} \\
 	&\quad + \frac{B-1}{\ln(2 - \frac{2}{3.4213})} 2^{-1 - \frac{2}{\ln 2}} \cdot \frac{2^{\frac{1}{U\ln2}}}{U} \cdot \frac{2^{\beta(1-L)}}{2^\beta - 1}.
\end{aligned}
\end{equation*}
This completes the proof of Equation~\eqref{eq20}.
\bibliographystyle{IEEEtran}
\bibliography{IEEEabrv,refs}
\end{document}